\documentclass[sn-chicago]{sn-jnl}

\jyear{2022}%

\theoremstyle{thmstyleone}%
\usepackage{setspace}

\theoremstyle{thmstyletwo}%

\theoremstyle{thmstylethree}%

\def\be{\begin{equation}}
\def\ee{\end{equation}}
\def\ben{\begin{eqnarray}}
\def\een{\end{eqnarray}}

\begin{document}

\title[Article Title]{Testing time evolution of mass distribution
of black hole mergers}


\author*[1]{\fnm{Teruaki} \sur{Suyama}}

\author[1]{\fnm{So} \sur{Okano}}


\affil*[1]{\orgdiv{Department of Physics}, \orgname{Tokyo Institute of Technology}, \orgaddress{\street{2-12-1 Ookayama}, \city{Meguro-ku}, \postcode{152-8551}, \state{Tokyo}, \country{Japan}}}


\abstract{
The detection of gravitational-wave events revealed that there are numerous populations of black hole (BH) 
binaries that can merge within the age of the Universe.
Although several formation channels of such binaries are known,
considerable theoretical uncertainties associated with each channel 
defeat the robust prediction of how much each channel contributes 
to the total merger rate density.
Given that the time evolution of the merger rate density in some channels is (exactly or nearly) independent of BH masses, clarifying 
this feature from observational data 
will shed some light on the nature of BH binaries.
On the basis of this motivation, we formulate a methodology to perform a 
statistical test of whether the mass distribution of BH mergers
evolves over time by hypothesis testing.
Our statistical test requires neither a priori specification of the mass distribution, 
which is largely uncertain, nor that of the time dependence of merger rate.
We then apply it to mock data for some concrete shapes of the 
merger rate density and show that the proposed method rejects/(does not reject)
the null hypothesis correctly for a large sample size.
After this verification, the method is applied to a catalog of the gravitational-wave events obtained during the LIGO-Virgo's
third observing run.
We find that the selection bias degrades the effectiveness
of our method for the O3 catalog owing to the reduction in the number of and the maximum redshifts of the merger events that we can explore.
Within the range where the method can be applied, there is no indication of the time evolution of the mass distribution of merger rate density.
This limitation will be eased in future observations containing more events, and our hypothesis testing
will help determine whether the merger rate density evolves over time independently of BH masses.
}




\maketitle

\section{Introduction}
The detection of gravitational waves from the mergers of black holes (BHs) has revealed
numerous populations of BH binaries in the Universe (\cite{LIGOScientific:2021usb, LIGOScientific:2021djp}).
Several formation channels have been proposed to explain the origin of such BH binaries
(e.g., see \cite{Mapelli:2021taw, Sasaki:2018dmp}).
Astrophysically, binary BHs can be directly formed as the end product of the stellar evolution of a field binary. 
As another formation channel, individual BHs formed in a dense environment can later form binaries dynamically. 
An exhaustive summary of the astrophysical scenarios for the formation of BH binaries
and their mergers as well as the expected merger rate in each scenario is found in
\cite{Mandel:2021smh}.
It is also possible that BHs that might have been created immediately after the Big-Bang (so-called primordial BHs (PBHs)) form binaries 
in the radiation dominated epoch and become the source of the detected GWs.
However, it is unclear if the current data favors the existence of PBHs (\cite{Franciolini:2021tla}).

Because of the considerable theoretical uncertainties in each channel,
it is still unknown how much each formation scenario 
contributes to merger rate (\cite{Belczynski:2021zaz}). 
Conversely, we may provide feedback to these theoretical models and update them
from observational data whose information in terms of merger rate, redshift, mass distribution,
spin, etc. has been increasing and will continue to increase in the future
owing to the progress of GW detectors.
Along this path, we can attempt to elucidate if
a single channel dominates the observed merger events or a few different channels 
nearly equally contribute.
To this end, we focus on the particular type of merger rate density written as
\be
\label{merger-rate}
{\cal R}(m_1,m_2,t)={\cal R}_0~ h(m_1,m_2) f(t).
\ee
Here, $m_1$ and $m_2$ are the masses of individual BHs in the binary measured 
in the source frame,
and $t$ is the cosmic age when the merger occurred.
$h(m_1,m_2)$ is normalized such that $\int h(m_1,m_2) dm_1 dm_2=1$, and
$f(t)$ is normalized such that $f(t_0)=1$, where $t_0$ is the age of the Universe.
Thus, ${\cal R}_0$ represents the merger rate at the present time.
The dimension of ${\cal R}$ is $/{\rm Gpc}^3 /{\rm yr}/M_\odot^2$,
and the rate density is defined for the comoving volume and cosmic time.
Thus, ${\cal R}(m_1,m_2,t) dV_c dt dm_1 dm_2$ represents the number of merger events of BHs with masses $m_1$ and $m_2$ 
which happen in the comoving volume $dV_c$ and during the time interval $(t,t+dt)$.

A crucial property of the merger rate density above is that it depends on the BH mass
and the merger time (i.e., redshift) in a separate manner:
it is simply given by the product of the mass-dependent function and the 
time-dependent function.
In other words, the mass distribution of merger rate density does not evolve over time.
Whether such evolution occurs depends on the formation channels.
In the isolated field binary scenario, 
the massive binary stars evolve into BH binaries after the mass transfer and the common envelope phase whose physical processes
have been investigated intensively (see e.g. \cite{Mapelli:2021taw} for a comprehensive review of this scenario).
In this case, the merger rate density is given by the convolution of star formation rate and the merger time delay distribution (\cite{Vitale:2018yhm}). 
Both of these may depend on the binary masses,
and the resultant merger rate density exhibits the time evolution
of the mass distribution (\cite{Dominik:2013tma, Tanikawa:2021qqi}).
In particular, the BH masses of the binary strongly depend on some factors such as the initial masses of the main sequence stars,
metallicity, whose typical values will change with redshift, and whether the pair-instability pulsation supernova occur (\cite{Belczynski:2016jno}). 
Dense environments such as globular clusters are the sites where BHs form binaries dynamically (\cite{Fragione:2018vty}), which undergo mergers,
and even successive mergers may form intermediate-mass BHs (\cite{Fragione:2021nhb}).
The merger rate of the BH binaries formed in the globular clusters is given by the convolution of 
the globular cluster formation rate and the merger time delay distribution (\cite{Rodriguez:2018rmd}). 
The time dependence of the mass distribution is determined by whether the mergers are dominated by
the binaries ejected from the globular cluster or
the binaries that remain inside until they merge.
In the former case, ejection efficiency depends on the BH masses,
which yields the time-dependent mass distribution (\cite{Rodriguez:2016kxx}). 
On the other hand, if the latter case is the dominant process, 
mergers follow shortly after a BH-BH encounter (\cite{Rodriguez:2018pss}) and 
the time evolution of the mass distribution will be suppressed (\cite{Samsing:2019dtb}).
As another merger channel, BH-BH encounters in galactic nuclei (\cite{Gondan:2017wzd, Rasskazov:2019gjw, Gondan:2020svr})
may be expected to have very little time evolution of their mass distribution (\cite{Chatterjee:2016thb, Yang:2020lhq}),
similarly to single-single GW captures in globular clusters (\cite{Samsing:2019dtb}). 
Young massive clusters and open clusters are also potential sites that contribute
to the GW events (\cite{Banerjee:2017mgr}). 
However, the evolution of merger rate density is not fully understood yet.
In addition to binary systems, BH mergers in triple or quadruple systems may have an important contribution to the GW events 
with some interesting observational consequences, such as a large eccentric orbit in the stage
corresponding to the frequency range
covered by ground interferometers, a large spin of the merged BHs, and formation of BHs in the low-mass gap range ($\lesssim 5~M_\odot$)
and the high-mass gap range ($\gtrsim 50~M_\odot$) (\cite{Antonini:2012ad, Fragione:2019hqt, Fragione:2020aki}).
More studies are needed to clarify how the merger rate density in such multiple systems evolves with redshifts. 
Finally, in the PBH scenario, the mass distribution
remains almost constant over time (\cite{Kocsis:2017yty, Raidal_2019}).

A quick overview of several representative scenarios of BH mergers above shows that each scenario suggests different features
of merger rate density.
If all (or some) of these scenarios predominantly contribute to the merger rate density, 
the total merger rate density becomes a superposition of the merger rate densities in individual scenarios and will not take
the separable form (\ref{merger-rate}) in general.
Thus, an observational confirmation of the time independence of the mass distribution
will disfavor the possibility that multiple scenarios contribute to merger events, and thus
the idea that a single channel is dominating the merger events is supported.
On the other hand, confirmation of the opposite case does not necessarily imply
the contribution of multiple channels since even a single channel may give a more complex merger rate density
than Eq.~(\ref{merger-rate}).
That is, if observations reveal the time evolution of the mass distribution,
more robust theoretical predictions concerning the merger rate density in individual scenarios are required to draw a reliable conclusion 
as to whether the merger events come from a single channel or multiple channels.

We have argued that the observational determination of the time evolution of the mass distribution provides an important key to
clarifying the origin of BH mergers.
Since there are currently non-negligible theoretical uncertainties of the merger rate density in each channel,
an agnostic statistical approach that is free from a priori assumptions on the shape of merger rate density would be a natural path to proceed,
which is a motivation for the study described in this paper. 
In light of this, our aim is to formulate a statistical method to 
test whether the observed merger rate density obeys 
the form of Eq.~(\ref{merger-rate}) \footnote{As for the total merger rate $\int {\cal R}dm_1 dm_2$,
there is a strong support for the increase toward the higher redshift (\cite{Fishbach:2021yvy, LIGOScientific:2021psn}).
See also the earlier work of \cite{Fishbach:2018edt} in which
the merger rate density taking the separable form with some specific functional shapes
was tested with the early LIGO-Virgo data.}.
As we will demonstrate, our method does not assume a priori the functional shapes
of $h (m_1,m_2)$ and $f(t)$, both of which strongly depend on 
the formation channel as well as the underlying assumptions that have considerable uncertainties due to the dearth of robust theoretical predictions of the merger rate density in the proposed channels.
Our approach is different from that in a previous study (\cite{Fishbach:2021yvy}) that also focused on the time evolution of the mass distribution
in that a Bayesian approach was taken in which some specific parametrizations for the merger rate density are assumed.
Thus, compared with the previous study, our approach is advantageous and new in that regard.
On the other hand, as we will discuss in Sec.~\ref{GWTC-2}, our approach cannot be straightforwardly applied when a selection bias,
which is an important factor for realistic data, is included.
In practice, only the merger events below some redshifts for which the selection bias does not significantly affect the sampling can be used, which degrades the
effectiveness of our approach owing to the reduction in data size as well as the decrease in the maximum redshift that we can explore.
Such an issue does not arise when using an Bayesian approach: the effect of the selection bias can be directly incorporated (\cite{Mandel:2018mve}).
In this way, our approach has both an advantage and a disadvantage compared with the Bayesian approach and plays a complementary role
in elucidating the origin of BH binaries.

\section{Formulation of the method}
For convenience, instead of the masses of individual BHs 
and the cosmic merger time,
we will use the total mass $M=m_1+m_2$, the mass ratio 
$q=\frac{m_2}{m_1} (m_2 \le m_1)$, and the redshift $z$ in the following analysis.
In terms of the new variables,
the expected number of merger events in the small mass area $dM dq$ and the redshift bin $(z,z+dz)$ during the observation time $T$ is given by
\be
\label{dN}
dN={\cal R}(M,q,z) \frac{T}{1+z} 
\frac{4\pi r^2(z) dz}{H(z)} \frac{M}{{(1+q)}^2}dMdq.
\ee
Here, $\frac{T}{1+z}$ is the time interval corresponding to $T$ in the source frame, $r(z)$ is the comoving distance to the redshift $z$,
$dV_c=\frac{4\pi r^2(z)}{H(z)} dz$ is the comoving volume of 
the thin shell $(z,z+dz)$,
and $\frac{M}{{(1+q)}^2}$ is the Jacobian due to the transformation
from $(m_1,m_2)$ to $(M,q)$.
Notice that the separability of the mass dependence and the merger time dependence,
which ${\cal R}$ possesses (i.e., Eq.~(\ref{merger-rate})),
is retained by $dN$, which plays a crucial role in the following analysis.

The number of events given above does not take into account the selection bias of the detector,
which becomes important for the region in the $(M,q,z)$ space close to and beyond the detection horizon.
This effect can be included by multiplying the detection probability $p_{\rm det}(M,q,z)$,
which is the probability that a given detector (or a network of detectors) detects a merger event with
masses $(M,q)$ occurring at $z$,
on the right-hand side of Eq.~(\ref{dN}).
The concrete shape of $p_{\rm det}$ depends on the detector (or a network of detectors) (\cite{Chen:2017wpg}).
Since $p_{\rm det}$ does not take the separable form in general,
the inclusion of the events corresponding to $p_{\rm det}<1$ invalidates the separability ansatz for $dN$.
In the following analysis, we assume an ideal case $p_{\rm det}=1$ or equivalently
consider only events much within the detection horizon. 
In Sec.~\ref{GWTC-2}, we briefly discuss how much the selection bias
affects the performance of our method.

\subsection{Basic idea}
\begin{figure}[t]
  \begin{center}
    \includegraphics[clip,width=10.0cm]{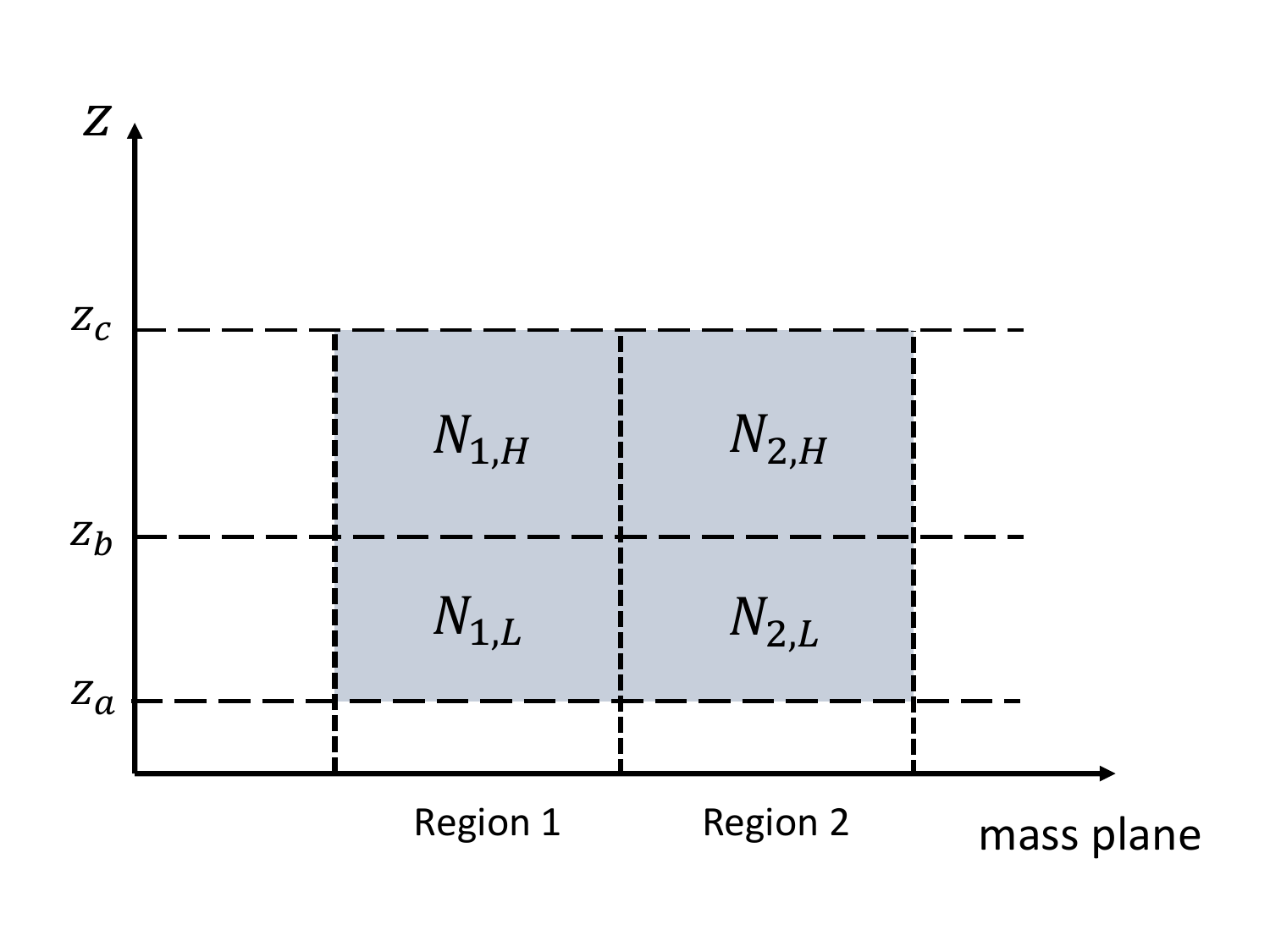}
    \caption{Definition of the division of the $(M,q,z)$ subspace into four regions. The horizontal axis
    represents the two-dimensional mass plane.}
    \label{fig2}
  \end{center}
\end{figure}

Let us take two distinct closed regions in the two-dimensional mass plane spanned
by $(M,q)$ (regions 1 and 2 in Fig.~\ref{fig2}) and two intervals $(z_a, z_b)$
indicated by $L$ and $(z_b,z_c)$ indicated by $H$ in the redshift axis.
The shapes of regions 1 and 2 are arbitrary.
For those regions, we can further define four regions as schematically described in Fig.~\ref{fig2}. 
For instance, ``$1,L$''stands for the region whose projection onto the mass plane coincides with
region 1 and the projection onto the redshift axis coincides with $(z_a,z_b)$.
Then, the expected number of merger events in this region is given by
\be
N_{1,L}=\int_{z_a}^{z_b} \int_{\rm region 1} dN. 
\ee
The expected number in the other regions can be expressed in a similar manner.
If the merger rate density takes the separable form (\ref{merger-rate}),
by substituting Eq.~(\ref{dN}), we obtain
\begin{align}
N_{1,L}=&\int_{z_a}^{z_b} \int_{\rm region 1} {\cal R}(M,q,z) \frac{T}{1+z} 
\frac{4\pi r^2(z) dz}{H(z)} \frac{M}{{(1+q)}^2}dMdq \nonumber \\
=&T {\cal R}_0 \bigg[ \int_{z_a}^{z_b} \frac{4\pi r^2(z)}{(1+z)}
\frac{f(z)}{H(z)}dz \bigg] \times \bigg[ 
\int_{\rm region 1} \frac{M}{{(1+q)}^2} h (M,q) dMdq \bigg].
\end{align}

It then follows that a ratio defined by
\be
\label{def-ratio}
R_A \equiv \frac{N_{A,H}}{N_{A,L}}=
\int_{z_b}^{z_c} \frac{4\pi r^2(z)}{(1+z)}
\frac{f(z)}{H(z)}dz \bigg/
 \int_{z_a}^{z_b} \frac{4\pi r^2(z)}{(1+z)}
\frac{f(z)}{H(z)}dz
\ee
becomes independent of A, where A stands for either 1 or 2.
Taking the contraposition of this statement, we can state that if the ratio $R_A$ depends on A,
the merger rate does not take the separable form similarly to Eq.~(\ref{merger-rate}).
Therefore, the hypothesis that the time dependence of the merger rate density is 
independent of the BH masses can be tested by checking whether the ratio $R_A$
is independent of A,
which is the basic idea underlying the following analysis.

\subsection{Hypothesis testing}
Having explained the basic idea, we formulate the statistical test
of whether the merger rate density takes the separable form given by Eq.~(\ref{merger-rate}).
We do this by hypothesis testing.
For technical convenience, instead of $R_A$, we will use a different quantity
defined by $p_A \equiv \frac{N_{A,H}}{N_{A,L}+N_{A,H}}=\frac{R_A}{1+R_A}$ in the following analysis.
Given that there is one-to-one correspondence between $p_A$ and $R_A$,
the use of $p_A$ is not essentially better than the use of $R_A$.
From the discussion in the previous subsection,
the merger rate density with the separable form leads to
the relation $p_1=p_2$. 
This is the mathematical expression that is suitable and ready
for making the statistical test that we want to conduct.
Since we aim to clarify if the time evolution of the merger rate density 
is independent of BH masses, we choose our null hypothesis $H_0$ to be
\be
H_0:~p_1=p_2
\ee
and the alternative hypothesis $H_1$ as
\be
H_1:~p_1 \neq p_2.
\ee
In what follows, we will explain how to test the hypothesis $H_0.$

We use the lower-case letter $n$ to denote the number of sample merger events in each subregion introduced in the previous subsection (see also Fig.~\ref{fig2}).
For instance, the number of events in region A ($A=1,2$) 
in $(z_a,z_b)$ is $n_{A,L}$ (the same for the others). 
Then, $n_{A,H}$ obeys the binomial distribution ${\rm Bin} (n_A, p_A)$,
where $n_A \equiv n_{A,L}+n_{A,H}$ is the sample size in region A.
For a large sample size, which is the assumption we are going to make, this distribution
is well approximated by the normal distribution, i.e., 
${\rm Bin} (n_A, p_A) \approx N(n_A p_A, n_A p_A (1-p_A))$.
Thus, a statistical quantity ${\bar p_A} \equiv \frac{n_{A,H}}{n_A}$ obeys the normal distribution
$N(p_A, p_A (1-p_A)/n_A)$.

Now, assuming that the hypothesis $H_0$ is true,
a test statistic $T_{\rm stat}$ defined by 
\be
\label{def-Tstat}
T_{\rm stat} \equiv \frac{{\bar p_1}-{\bar p_2}}{\sqrt{{\bar p}(1-{\bar p}) \left( \frac{1}{n_1}+
\frac{1}{n_2}\right) }},
\ee
where ${\bar p} \equiv \frac{n_1 {\bar p_1}+n_2 {\bar p_2}}{n_1+n_2}$ is the pooled population proportion,
obeys the normal distribution $N(0,1)$.
Thus, we can/(cannot) reject the hypothesis $H_0$ at a significance level $\alpha$
by computing whether the magnitude of $T_{\rm stat}$ is larger/smaller than $\sqrt{2} {\rm Erfc}^{-1} (\alpha)$ (two-tailed test),
where ${\rm Erfc}(x) \equiv \frac{2}{\sqrt{\pi}} \int_x^\infty e^{-t^2}dt$ is the complementary error function.
This is the main strategy of our statistical test.

To have a rough idea of how much the above-mentioned method works 
for testing the merger rate density given by Eq.~(\ref{merger-rate}),
let us crudely estimate the required sample size to reject the hypothesis $H_0$ at 
the $5\%$ significance level for
the case where the merger rate density does not take the separable form. 
To this end, we perform a simple parametrization for such a case using $p_1-p_2=\Delta p ~(\neq 0)$.
The factor in the denominator $\sqrt{{\bar p}(1-{\bar p})}$
takes the maximum at ${\bar p}=\frac{1}{2}$,
and we choose the value $\sqrt{{\bar p}(1-{\bar p})}=\frac{1}{2}$
that minimizes $T_{\rm stat}$ when other parameters are fixed.
By replacing ${\bar p_1}-{\bar p_2}$ appearing in the numerator of Eq.~(\ref{def-Tstat})
by $\Delta p$ as a representative value,
we reject the hypothesis $H_0$ when $\| \Delta p \| > 0.98 \sqrt{\frac{1}{n_1}+\frac{1}{n_2}}$.  
The right-hand side of this condition becomes minimum at $n_1=n_2$ for a fixed $n=n_1+n_2$.
Thus, the minimum sample size $n$ needed to reject $H_0$ for the merger rate density 
parametrized by $\Delta p$ is at least about $3.84/{(\Delta p)}^2$.

\section{Demonstration}
\label{demonstration}
In this section, we demonstrate the statistical approach introduced
in the previous section by studying the distribution of $T_{\rm stat}$ of 
the samples for some specific merger rate densities.
By doing this, we can determine the possible effectiveness of the proposed method when it is applied to future observational data. 
In what follows, we study two representative cases for the merger rate density: 
the separable form and the nonseparable form.
The study of the former case allows us to confirm the robustness of the method 
by checking that the distribution of $T_{\rm stat}$ of the mock data
with a large sample size
approximates the normal distribution $N(0,1)$.
This can also be used to check the probability of making a type I error
for the null hypothesis $H_0$.
Meanwhile, the analysis of the latter case illustrates how efficiently
the null hypothesis $H_0$ is rejected when the alternative hypothesis $H_1$
is true.
Thus, this case provides us a good estimate of making a type II error.

As it is evident from how the statistical method is formulated, 
the choice of the shapes of regions 1 and 2 is completely arbitrary.
In the following analysis, we define these regions as
\begin{align}
\nonumber
{\rm region~1}=\{ (M,q) \| M \le M_{\rm div}, ~0\le q \le 1 \}, \\
{\rm region~2}=\{ (M,q) \| M \ge M_{\rm div}, ~0\le q \le 1 \}.
\end{align}
Here, $M_{\rm div}$ is the critical total mass that divides regions 1 and 2.

\subsection{Separable form}
In this subsection, we study the merger rate density that takes the
separable form (\ref{merger-rate}).
As the shape of the merger rate density, we consider two examples:
the mergers of the PBH binaries and the mergers of the astrophysical
BH binaries that follow the star formation rate.

\subsubsection{PBH mergers}
\label{PBH-mergers}
As for the PBH mergers, we assume that 
the mass dependent part $h (m_1,m_2)$ is given by
\be
\label{PBH-R0}
h (m_1, m_2)=C \psi (m_1) \psi (m_2),
\ee
where $\psi (m)$ is the PBH mass function, and $C$ is a normalization constant such that $\int h(m_1,m_2) dm_1 dm_2=1$.
The time evolution part $f(t)$ is given by
\be
f(t) ={\left( \frac{t}{t_0} \right)}^{-\frac{34}{37}},
\ee
where $t_0$ is the age of the Universe.
This time dependence is realized for PBH binaries that formed in the radiation- 
dominated epoch (\cite{Nakamura:1997sm, Ioka:1998nz, Sasaki:2016jop}), and this formation channel dominates the PBH merger rate 
if the PBH binaries are not disrupted by other gravitational sources throughout 
their subsequent evolution (\cite{Sasaki:2018dmp}).
The PBH mass function $\psi (m)$ strongly depends on the models
of the early universe.
In this paper, we consider the log-normal shape 
\be
\psi (m)= \exp \left( -\frac{1}{2\sigma^2}\ln^2 \left( \frac{m}{m_0} \right) \right),
\ee
which is a widely used phenomenological functional form (\cite{Carr:2017jsz}).
Here, $m_0$ and $\sigma$ are free parameters, and we choose
$M_0=40~M_\odot, \sigma=0.2$ in our analysis.

\begin{figure}
\begin{center}
\begin{tabular}{cc}
\begin{minipage}[t]{0.5\hsize}
\includegraphics[width=6cm]{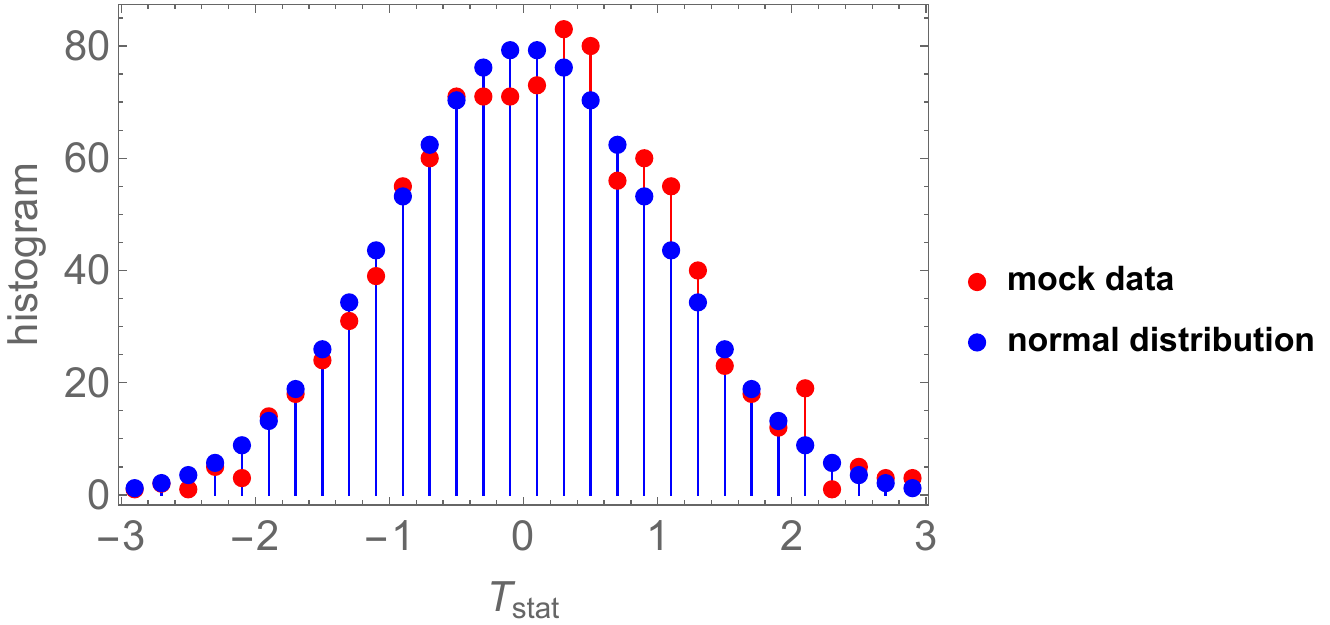}
\end{minipage}
\begin{minipage}[t]{0.5\hsize}
\includegraphics[width=6cm]{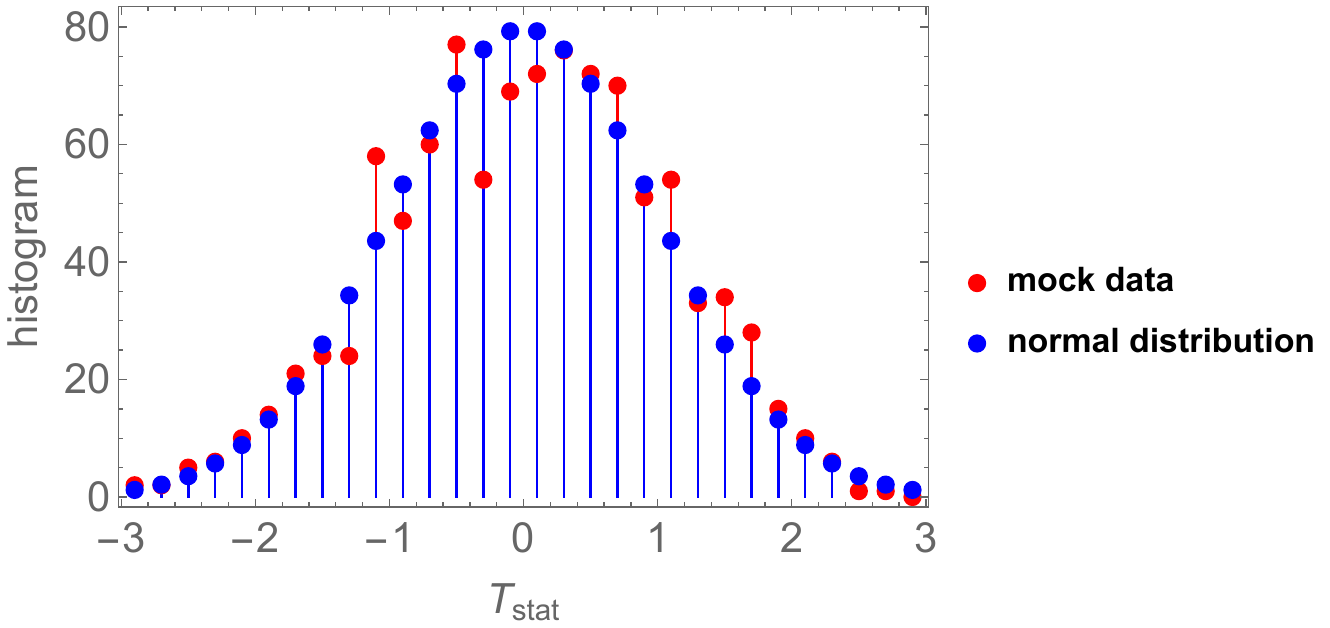}
\end{minipage}
\end{tabular}
\caption{Histograms of $T_{\rm stat}$ of one thousand realizations for the 
merger rate density with the separable form.
Each realization has $n_1+n_2=1000$ sample size.
The left panel is for the PBH merger rate density,
and the right panel is for the merger rate density of the astrophysical BHs.
The explicit shape of the merger rate density and the underlying assumptions for each model are given in the main text.}
\label{separable-Tstat}
\end{center}
\end{figure}

The red dots in the left panel of Fig.~\ref{separable-Tstat} are the plots in the histogram
of $T_{\rm stat}$ of one thousand realizations, each of which
has $n_1+n_2=1000$ sample size. 
The blue dots are the histogram of $T_{\rm stat}$ obeying the normal
distribution $N(0,1)$ that, as discussed in the previous section, 
should be realized if the merger rate density takes the separable form.
As evident from the figure, the distribution of $T_{\rm stat}$ of
the sample data is consistent with the normal distribution,
which explicitly demonstrates the validity of the statistical method 
presented in the previous section.
This is justified quantitatively by computing the $p$-value based on the Anderson-Darling test, which yields $p=0.82$.
Our choice of parameters defining the four regions shown in Fig.~\ref{fig2} is $(M_{\rm div}, z_b, z_c)=(80~M_\odot, ~0.5, ~1.0)$.

\subsubsection{Mergers of astrophysical BHs}
\label{separable-abh}
As for the shape of the merger rate density of the astrophysical BHs,
we adopt the simple phenomenological model studied by \cite{LIGOScientific:2018jsj}
\footnote{This model is called {\it model B} in \cite{LIGOScientific:2018jsj}}.
This model is disfavored by the updated analysis presented by
\cite{LIGOScientific:2020kqk}.
Nevertheless, we adopt this model in our analysis because our purpose
is to demonstrate the effectiveness of our statistical method, and 
the use of the simple model would be sufficient for this purpose.
Notice also that there remains a possibility that only a fraction
of the merger events obey this model.
The mass-dependent part $h(m_1,m_2)$ in this model is given by
\be
\label{R0-astroBH}
h(m_1,m_2)=C m_1^{-\alpha} {\left( \frac{m_2}{m_1} \right)}^{\beta_q}
\Theta (m_2-m_{\rm min}) \Theta (m_{\rm max}-m_1) \Theta (m_1-m_2),
\ee
where $\Theta (x)$ is the Heaviside function
and $C$ is the normalization constant.
This shape contains four free parameters: $\alpha, \beta_q, m_{\rm min}$, 
and $m_{\rm max}$. 
In the analysis by \cite{LIGOScientific:2018jsj}, this model has been compared with the 
data obtained during the first and second observation runs of
LIGO and Virgo, and the posteriors of the four free parameters are derived. 
In our analysis, we choose them to be $\alpha=1.3,~\beta_q=7$,
and $(m_{\rm min}, ~m_{\rm max})=(8~M_\odot,~40~M_\odot)$,
which are consistent with the posteriors mentioned above.
The time evolution part $f(t)$ is assumed to exactly follow the star formation rate (\cite{Madau:2014bja}), namely,
\be
\label{f-astroBH}
f(z)=\frac{1}{0.997}\frac{{(1+z)}^{2.7}}{1+{\left( \frac{1+z}{2.9} \right)}^{5.6}}.
\ee
Here, we abuse the notation of $f(t)$ by changing the argument from the cosmic
time $t$ to the redshift $z$ since the star formation rate is commonly
given in terms of $z$.

The red dots in the right panel of Fig.~\ref{separable-Tstat} show the 
histogram of $T_{\rm stat}$ of one thousand realizations for the
model defined by Eqs.~(\ref{R0-astroBH}) and (\ref{f-astroBH}).
The sample size of each realization is the same as that in the case of the PBH mergers
(i.e., the left panel); $n=n_1+n_2=1000$. 
Our choice of parameters defining the four regions shown in Fig.~\ref{fig2} is $(M_{\rm div}, z_b, z_c)=(60~M_\odot, ~0.5, ~1.0)$.
The blue dots are the histogram of $T_{\rm stat}$ obeying the normal
distribution $N(0,1)$.
As it is the same as that in the left panel, the distribution of $T_{\rm stat}$ of
the sample data is consistent with the normal distribution ($p$-value based on the Anderson-Darling test is $0.06$).
Thus, from the two examples, we are able to confirm that the hypothesis testing
can reject the null hypothesis $H_0$ at a given significance level $\alpha$ if 
the value of $T_{\rm stat}$ constructed from data is larger than 
$2{\rm Erfc}^{-1} (\alpha)$.

\subsection{Non-separable form}
Having checked that the probability of rejecting the null hypothesis $H_0$ 
even when $H_0$ is true is controlled by the significance level,
we next investigate how likely it is not to reject the null hypothesis
even when it is false.

\subsubsection{Case 1: Toy model}
\label{non-s:toy-model}
To this end, we first consider an extreme toy model in which the merger rate
density is given by
\be
\label{non-separable-1}
{\cal R}(m_1,m_2,z)={\cal R}_0 h(m_1,m_2)
\left( \Theta (M_c-M)+{(1+z)}^5 \Theta (M-M_c) \right),
\ee
where $h(m_1,m_2)$ is defined by Eq.~(\ref{R0-astroBH}) and $M_c$ is a free parameter that we choose as $M_c=40~M_\odot$.
Whereas the merger rate of the BH binaries with $m_1+m_2 <M_c$ does not
evolve with the redshift, 
that with $m_1+m_2 >M_c$ has a strong dependence on the redshift 
as $\propto {(1+z)}^5$.
Thus, the merger rate density (\ref{non-separable-1}) takes the
non-separable form that does not belong to the class defined by
Eq.~(\ref{merger-rate}) and provides one example of the alternative
hypothesis $H_1$.

\begin{figure}[t]
  \begin{center}
    \includegraphics[clip,width=7.0cm]{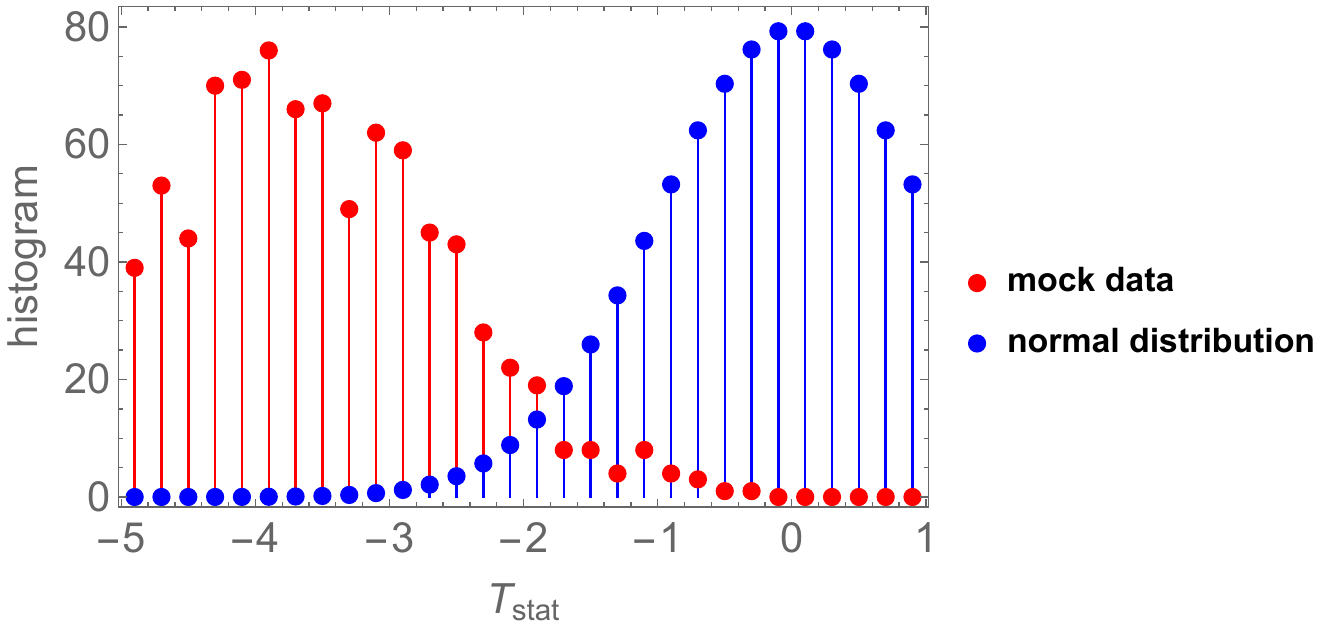}
    \caption{Histograms of $T_{\rm stat}$ of one thousand realizations for the 
merger rate density with the specific non-separable given by Eq.~(\ref{non-separable-1}).
Each realization has $n_1+n_2=1000$ sample size.}
    \label{mixed-case1}
  \end{center}
\end{figure}

The red dots in the left panel of Fig.~\ref{mixed-case1} show the 
histogram of $T_{\rm stat}$ of one thousand realizations of
the merger rate density given by Eq.~(\ref{non-separable-1}).
It is clear that the distribution of $T_{\rm stat}$ is markedly shifted to the negative side
and peaks at around $T_{\rm stat} =-4$. 
Our choice of parameters defining the four regions shown in Fig.~\ref{fig2} is $(M_{\rm div}, z_b, z_c)=(40~M_\odot, ~0.7, ~1.0)$.
For this choice, $p_1$ and $p_2$ are found to be $(p_1,~p_2)=(0.51,~0.74)$ and
\be
\label{def-c1}
\int_{\rm Region 1} {\cal R}(m_1,m_2, t) dm_1 dm_2 dt~
= c_1 \int_{\rm Region 1+Region 2} {\cal R}(m_1,m_2, t) dm_1 dm_2 dt
\ee
with $c_1 \approx 0.055$.
Using these values as typical ones for the quantities appearing 
in the definition of $T_{\rm stat}$ (\ref{def-Tstat}), we can estimate the
typical value of $T_{\rm stat}$ as
\be
\label{coefficient-Tstat}
T_{\rm stat} = -3.8~ \sqrt{ \frac{n}{1000}},
\ee
which is consistent with the peak value of the mock data in Fig.~\ref{mixed-case1}.

For the current example, we find that the number of realizations yielding $T_{\rm stat}>-2$
out of our particular one thousand realizations is 45.
Thus, if the real merger rate density is given by Eq.~(\ref{non-separable-1}),
we can reject the null hypothesis $H_0$ at about the $5\%$ significance level for the adopted parameter values 
when the sample size is larger than $1000$.

\begin{figure}
\begin{center}
\begin{tabular}{cc}
\begin{minipage}[t]{0.5\hsize}
\includegraphics[width=5cm]{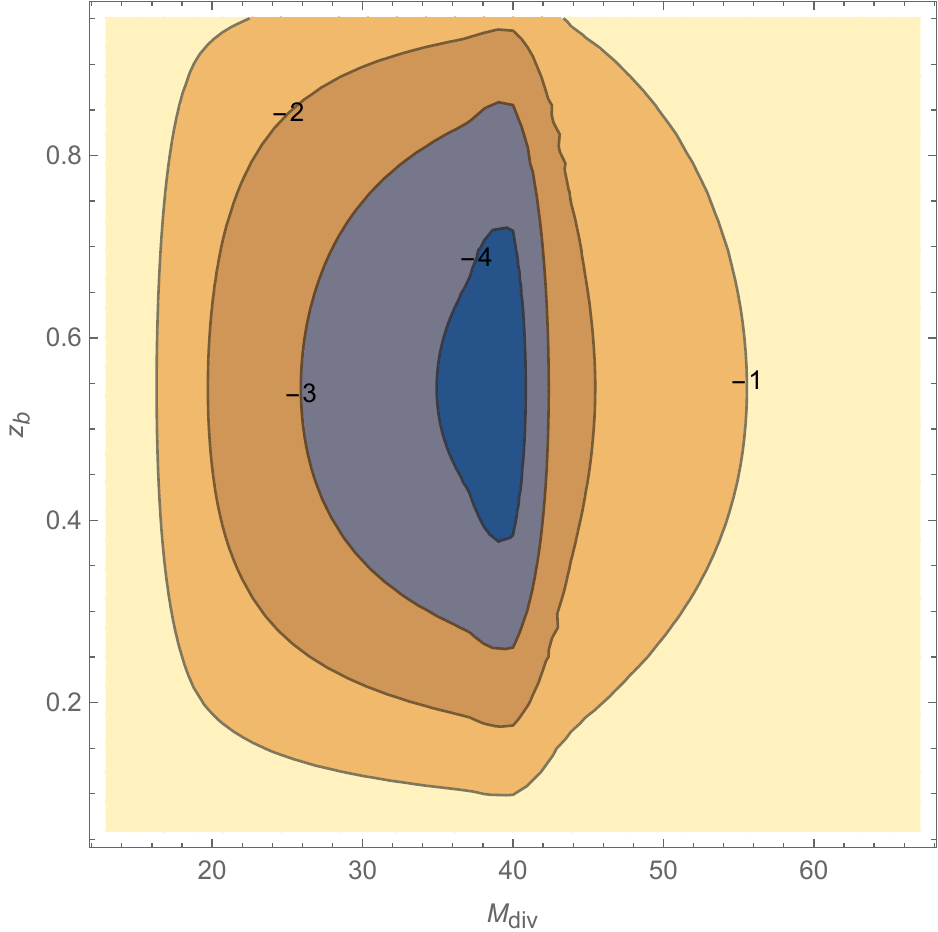}
\end{minipage}
\begin{minipage}[t]{0.5\hsize}
\includegraphics[width=5cm]{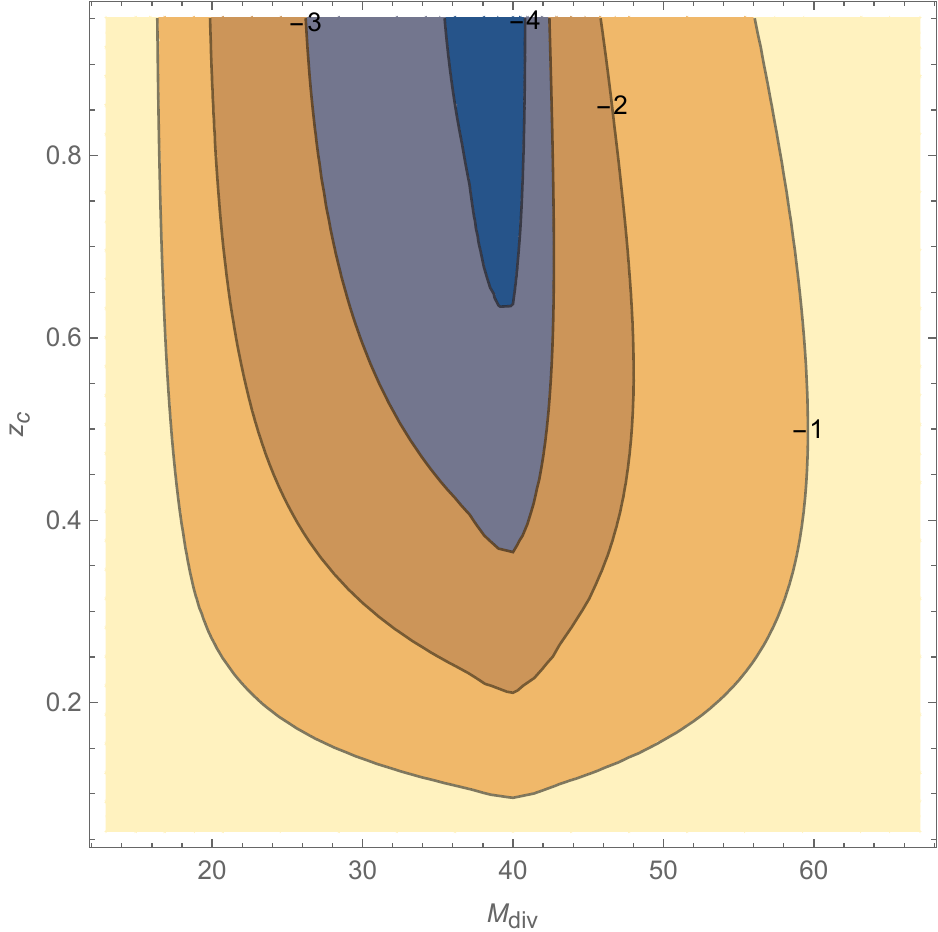}
\end{minipage}
\end{tabular}
\caption{Contour plot of $T_{\rm stat}$ for $n=1000$.
In the left panel, $M_{\rm div}$ and $z_b$ are varied while $z_c=1$ is fixed.
In the right panel, $M_{\rm div}$ and $z_c$ are varied while $z_b$ is fixed to $z_c/2$.}
\label{contour-Tstat}
\end{center}
\end{figure}

The width of the distribution of the red dots in Fig.~\ref{mixed-case1} is ${\cal O}(1)$.
Actually, this does not depend on $n$ because the typical variation of $T_{\rm stat}$
due to the randomness of the sampling gives a scaling 
$\frac{\delta T_{\rm stat}}{T_{\rm stat}} \propto n^{-1/2}$,
and combining it with the scaling $T_{\rm stat} \propto n^{1/2}$ shown in the above equation 
yields $\delta T_{\rm stat} \propto n^0$, whereas the
numerical value of the proportionality coefficient, which is ${\cal O}(1)$, 
may depend on the underlying merger rate density as well as the
parameters $(M_{\rm div}, z_b, z_c)$.
To see the latter point in more detail, Fig.~\ref{contour-Tstat} shows the contour of the
coefficient of $\sqrt{n/1000}$ of Eq.~(\ref{coefficient-Tstat}) in the $(M_{\rm div}, z_b)$ plane ($z_c=1$) in the left panel
and in the $(M_{\rm div}, z_c)$ plane ($z_b=z_c/2$) in the right panel.
From the left panel, we clearly see that $-T_{\rm stat}$ peaks at around $M_{\rm div}=40~M_\odot$.
This is natural because it corresponds to the boundary $M_c$ in the merger rate density (\ref{non-separable-1}), which gives the different redshift evolutions:
the difference in the distribution of the merger events between regions 1 and 2 becomes the most prominent when
the different redshift evolutions (i.e., the first and second terms in Eq.~(\ref{non-separable-1})) are separately covered by different regions.
This consideratation is corroborated by the right panel where $-T_{\rm stat}$ decreases as $z_c$ decreases even when $M_{\rm div}$ is
fixed to $40~M_\odot$.
By restricting the region of the merger events to low redshifts only, we find that the first and second terms are nearly identical
and, consequently, the distribution of the merger events in region 1 becomes indistinguishable from that in region 2.
From this investigation, we find that the effectiveness of the current method is controlled by three factors:
the total number of merger events $n$, the maximum redshift covered by observations $z_c$, and the characteristic BH mass providing 
a non-separable measure in the merger rate density.
Although the first two are solely determined by observations, the last one depends on the concrete shape of the 
underlying merger rate density, which we do not know a priori in real observations.
When applying our method to real data, we will need to compute $T_{\rm stat}$ and test the null hypothesis 
for various values of $M_{\rm div}$.

\subsubsection{Case 2: Mixture of astrophysical BHs and PBHs}
The above example is unrealistic in the sense that it is not based on astrophysics
and is introduced only for the purpose of demonstrating the principle of our statistical method.
In the second example, we consider a less extreme case in which the merger rate is a mixture of the
mergers of the astrophysical BHs and those of PBHs, each of which has been separately investigated in the 
previous subsection. Namely, we assume the merger rate density given by
\be
\label{R-mixed-case2}
{\cal R}(m_1,m_2,t)=(1-r) {\cal R}_{\rm astro}(m_1,m_2,t)+r {\cal R}_{\rm PBH}(m_1,m_2,t),
\ee
where ${\cal R}_{\rm astro}$ and ${\cal R}_{\rm PBH}$ are the merger rate densities of the astrophysical
BHs and PBHs introduced by Eqs.~(\ref{PBH-R0})-({\ref{f-astroBH}}), respectively.
Here, we choose the normalization of the individual contributions such that
they give the same merger rate at the present time $t_0$;
\be
\int {\cal R}_{\rm astro} (m_1,m_2,t_0) dm_1 dm_2
=\int {\cal R}_{\rm PBH} (m_1,m_2,t_0) dm_1 dm_2.
\ee
Thus, $r$ denotes the fraction of the PBH contribution to the total merger rate at the present time.
Since ${\cal R}_{\rm astro}$ and ${\cal R}_{\rm PBH}$ have different $z$ 
dependences, the above merger rate density is non-separable for $0<r<1$.

\begin{figure}
\begin{center}
\begin{tabular}{cc}
\begin{minipage}[t]{0.5\hsize}
\includegraphics[width=5cm]{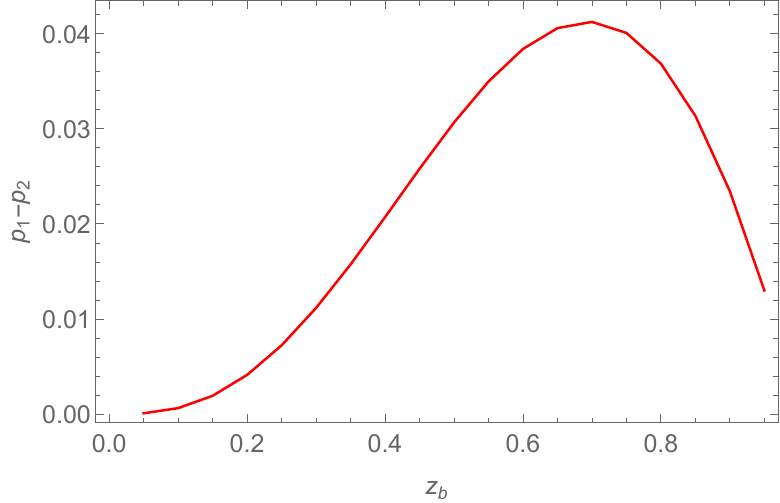}
\end{minipage}
\begin{minipage}[t]{0.5\hsize}
\includegraphics[width=6cm]{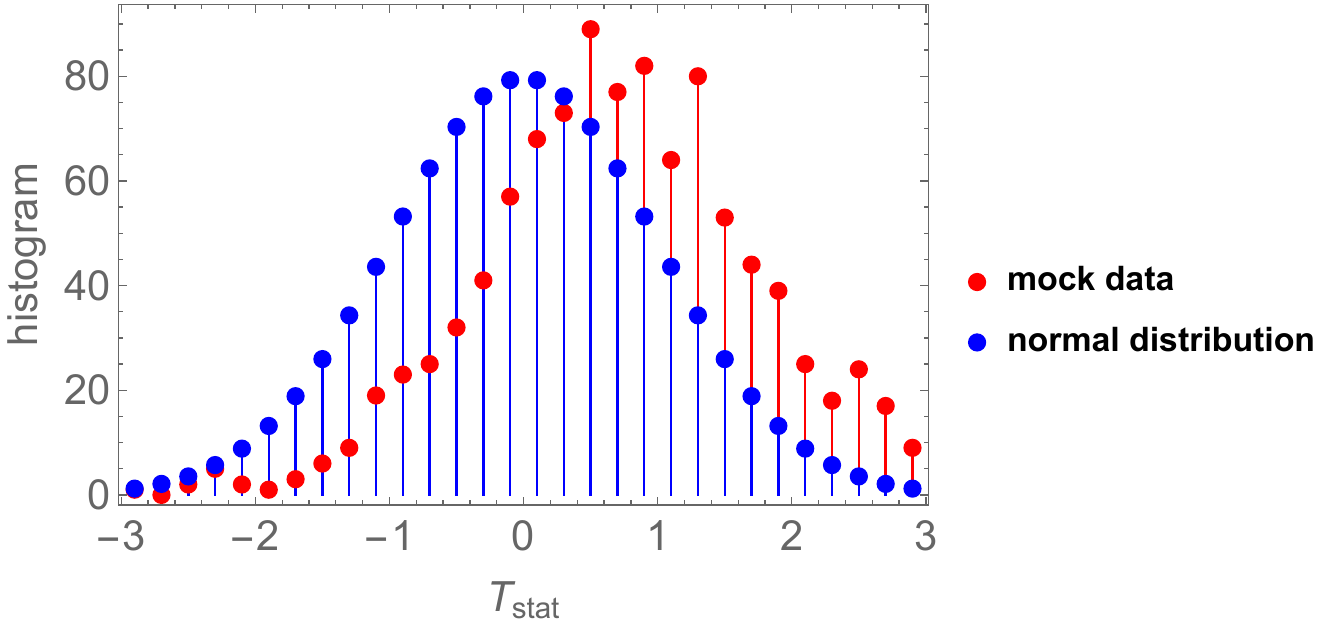}
\end{minipage}
\end{tabular}
\caption{
Left panel showing $p_1-p_2$ as a function of $z_b$ with other parameters ($M_{\rm div}, z_c$)
being fixed for the merger rate density given by Eq.~(\ref{R-mixed-case2}).
Right panel showing the histogram of $T_{\rm stat}$ of one thousand realizations for the 
same merger rate density as that in the left panel.
Each realization has $n_1+n_2=1000$ sample size.
}
\label{mixed-case2}
\end{center}
\end{figure}

Our choice of parameters $M_{\rm div}$ and $z_c$ is $60~M_\odot$ and $1.0$.
The left panel of Fig.~\ref{mixed-case2} shows $p_1-p_2$ as a function of $z_b$.
From this, we find that $p_1-p_2$ becomes maximal at $z_b \approx 0.7$.
In the following analysis, we take $z_b =0.7$.
The right panel of Fig.~\ref{mixed-case2} shows the histogram
of $T_{\rm stat}$ of the mock data of the $1000$ sample size for the merger rate density given by Eq.~(\ref{R-mixed-case2}).
The result shows that the peak of the histogram is located at about $T_{\rm stat}=1.5$.
For the current merger rate density with the same values of the parameters 
as those adopted for generating the mock data,
we find $(p_1, p_2) \approx (0.636, 0.595)$ and
\be
\int_{\rm Region 1} {\cal R}(m_1,m_2, t) dm_1 dm_2 dt~
= c_1 \int_{\rm Region 1+Region 2} {\cal R}(m_1,m_2, t) dm_1 dm_2 dt
\ee
with $c_1 \approx 0.54$.
Using these average values as typical ones for the quantities appearing 
in the definition of $T_{\rm stat}$ (\ref{def-Tstat}), we can estimate the
typical value of $T_{\rm stat}$ as
\be
\label{Tstat-n}
T_{\rm stat} = 1.3~ \sqrt{ \frac{n}{1000}}
\ee
in terms of the sample size $n=n_1+n_2$.
As expected, this value is consistent with the peak value of $T_{\rm stat}$ 
at which the histogram of $T_{\rm stat}$ of the mock data becomes maximal,
and the width of the distribution is ${\cal O}(1)$.

To summarize, these examples demonstrate that if a given non-separable merger
rate density is realized in nature, the typical value of $T_{\rm stat}$
given by
\be
\frac{p_1-p_2}{\sqrt{p(1-p) \left( \frac{1}{c_1}+
\frac{1}{c_2}\right) }} \sqrt{n},
\ee
where $p = c_1 p_1+c_2 p_2$ and $c_1$ is defined by Eq.~(\ref{def-c1}) and $c_2 \equiv 1-c_1$,
provides a good indicator of whether the null hypothesis $H_0$ can be rejected
for the data containing $n$ merger events.

\subsection{Effect of measurement error of source parameters on the distribution of $T_{\rm stat}$}
Thus far, in our analysis, we assumed no observational errors on the parameters 
of the source binaries $(M, q, z)$.
In reality, those parameters are always accompanied by errors.
Since such errors will let us mistakenly place the position of the
individual merger event in the $(M,q,z)$ space, 
the determination of the number of merger events in each region described in Fig.~\ref{fig2} is affected accordingly.
As a result, it is expected that the effectiveness of the hypothesis testing will be degraded to some extent.
In this subsection, we evaluate the significance of the effect
of the errors on the hypothesis testing.

To simplify our analysis, we assign $10\%$ error randomly to the three parameters $(M,q,z)$ of any merger events. 
This is not true in reality since the magnitude of the error in general
depends on the binary masses and the distance to the binary.
However, the following analysis based on this simplification enables us
to capture how the observational error affects our method at least qualitatively.

As an explicit example, we first consider the merger rate density
of the PBH binaries investigated in Sec.~\ref{PBH-mergers} with the same values of the parameters.
For the binary parameters $(M,q,z)$ of each 
randomly generated merger event, we multiply a
random number corresponding to the $10\%$ error,
namely, we change the parameters $(M_i, q_i, z_i)$ of the $i$-th merger event
into $(M_i (1+a_i), q_i (1+b_i), z_i (1+c_i))$,
where $(a_i, b_i, c_i)$ are uncorrelated random numbers in the range $[-0.1,0.1]$.

\begin{figure}[t]
  \begin{center}
    \includegraphics[clip,width=9.0cm]{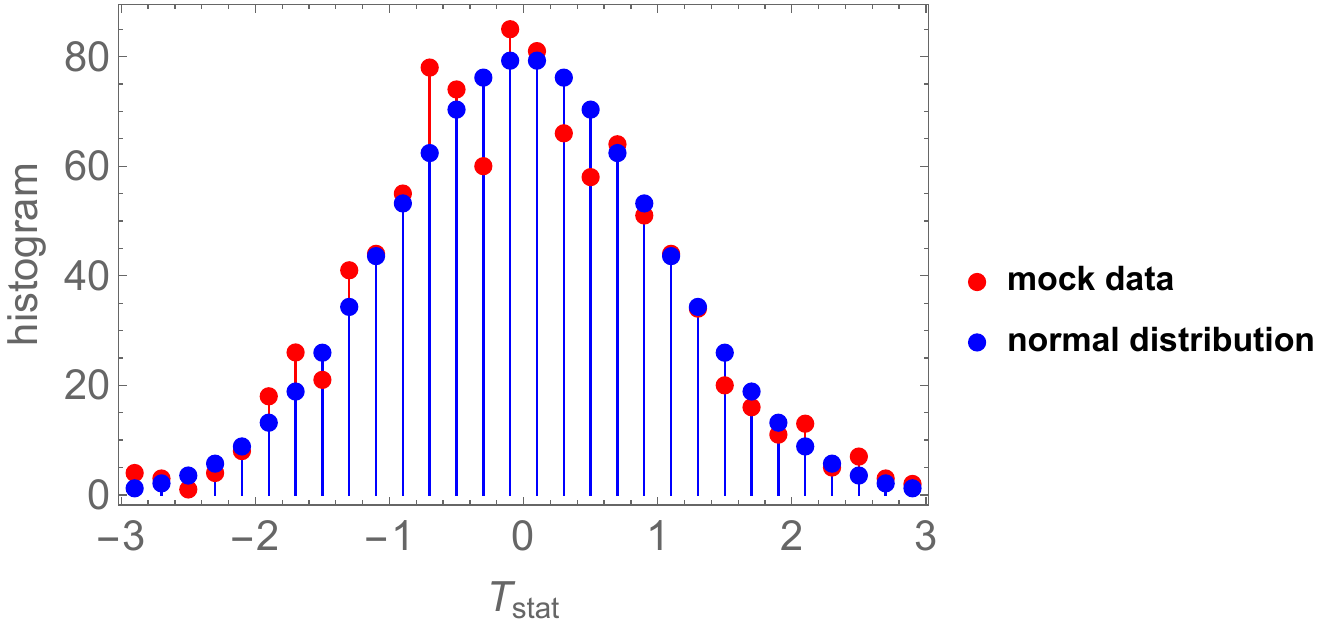}
    \caption{Histogram of $T_{\rm stat}$ of one thousand realizations for the 
merger rate density of the PBH binaries with the observational
errors ($10\%$) of the binary parameters being included.
Each realization has $n_1+n_2=1000$ sample size.}
    \label{Tstat-error-included}
  \end{center}
\end{figure}

\begin{figure}[t]
\begin{center}
\begin{tabular}{cc}
\begin{minipage}[t]{0.5\hsize}
\includegraphics[width=6cm]{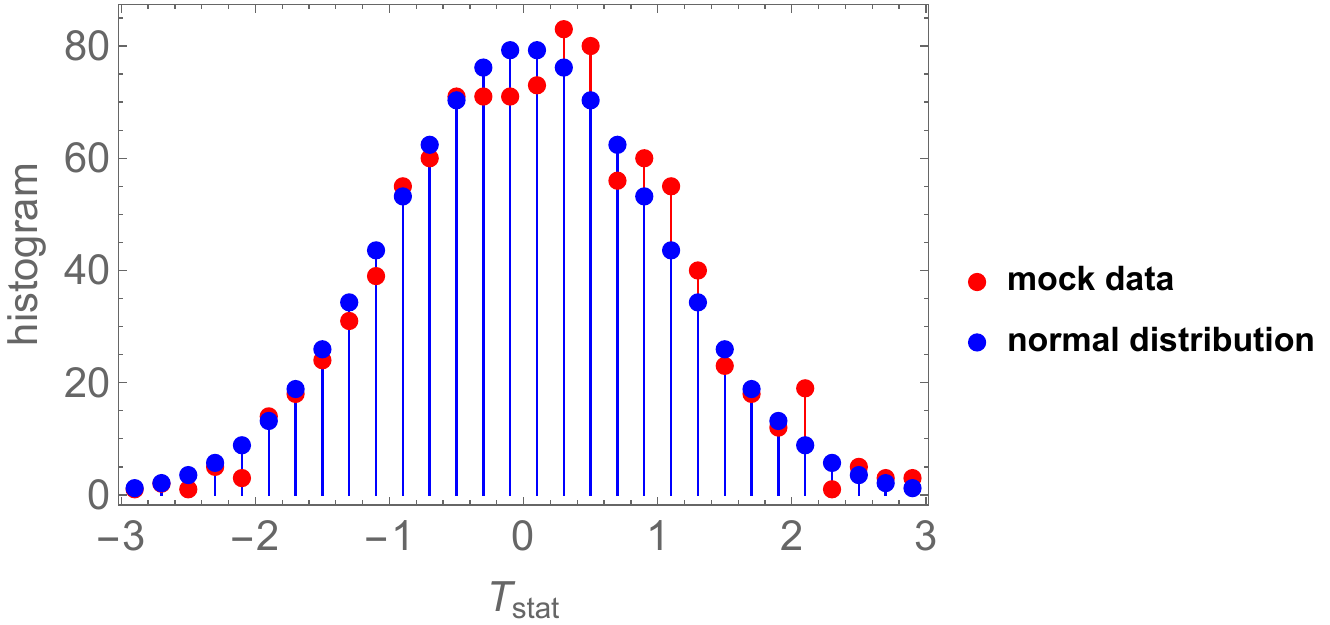}
\end{minipage}
\begin{minipage}[t]{0.5\hsize}
\includegraphics[width=6cm]{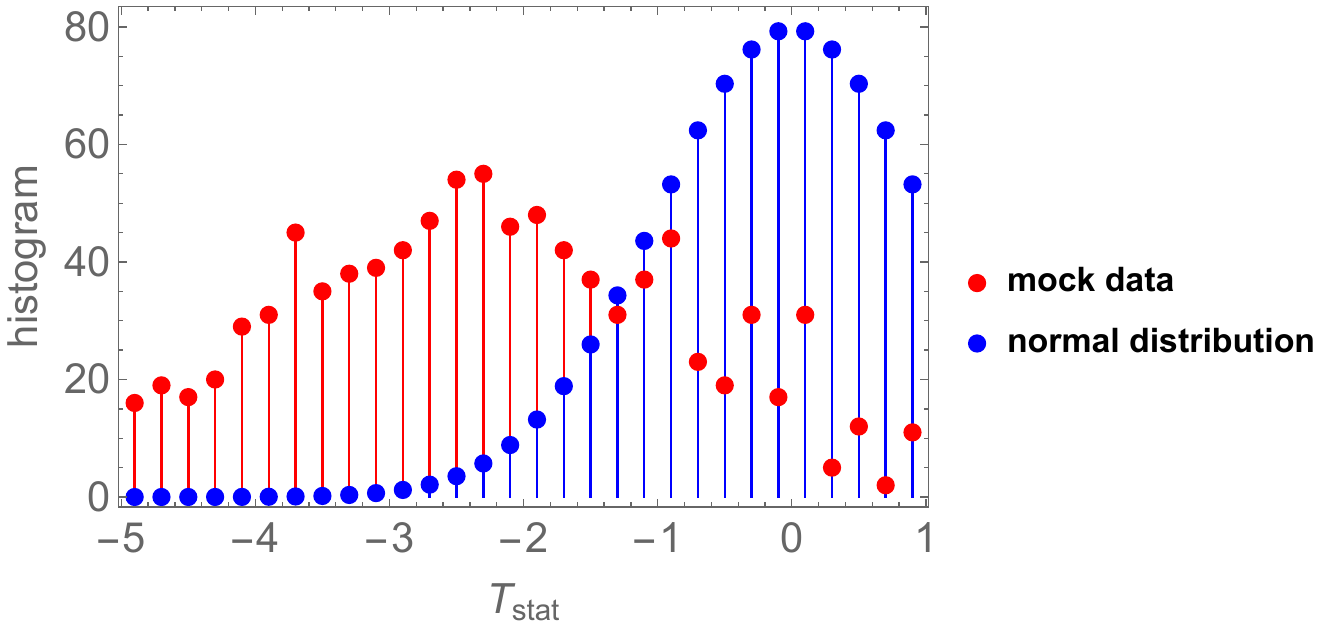}
\end{minipage}
\end{tabular}
\caption{
Left panel: histogram of $T_{\rm stat}$ of one thousand realizations for the 
merger rate density of the PBH binaries with the observational
errors ($50\%$) of the binary parameters being included.
Right panel: histogram of $T_{\rm stat}$ of one thousand realizations for the 
merger rate density considered in Sec.~\ref{non-s:toy-model} with the observational
errors ($10\%$) of the binary parameters being included.
}
\label{error-case2}
\end{center}
\end{figure}

Fig.~\ref{Tstat-error-included} shows the histogram of $T_{\rm stat}$
of one thousand realizations, each of which contains $n_1+n_2=1000$ sample size. 
As we can see, the histogram of $T_{\rm stat}$ of the mock data is hardly distinguishable from
that of the normal distribution.
Namely, the observational errors with the current magnitude 
minimally affect the effectiveness of the hypothesis testing. 
This result may be understood as follows.
The observational error, by which some events near the boundaries dividing the four subspaces in Fig.~\ref{fig2}
are counted randomly in different subspaces, erases the contrast among the number of events in each subspace. 
As a result, $p_1$ and $p_2$ tend to take similar values, which suppress $T_{\rm stat}$.
That is, the observational error should effectively make the apparent mass distribution of the merger events 
look more independent of the redshift.
To corroborate this explanation, we also constructed the histogram of $T_{\rm stat}$ where the error has now been increased
to $50\%$, which is shown as the left panel of Fig.~\ref{error-case2}.
As we can verify, the histogram is still consistent with the normal distribution $N(0,1)$.
As another example, the right panel of Fig.~\ref{error-case2} shows the histogram of $T_{\rm stat}$
of the merger rate density considered in Sec.~\ref{non-s:toy-model} with $10\%$ errors added.
Clearly, the histogram shifts toward the normal distribution $N(0,1)$ compared with those in Fig.~\ref{mixed-case1}
for which the observational error is not included.
Actually, the probability that $T_{\rm stat}$ is within $95\%$ region of the normal distribution is fairly larger than $5\%$.

These investigations show that the inclusion of the observational error tends to favor the null hypothesis
compared with the case where no errors are included.
Thus, if the null hypothesis is rejected even after including the observational errors, 
it is a strong indication that the mass distribution of the merger events evolves with redshift.

\subsection{Application to O3 data}
\label{GWTC-2}
\begin{figure}[t]
  \begin{center}
    \includegraphics[clip,width=7.0cm]{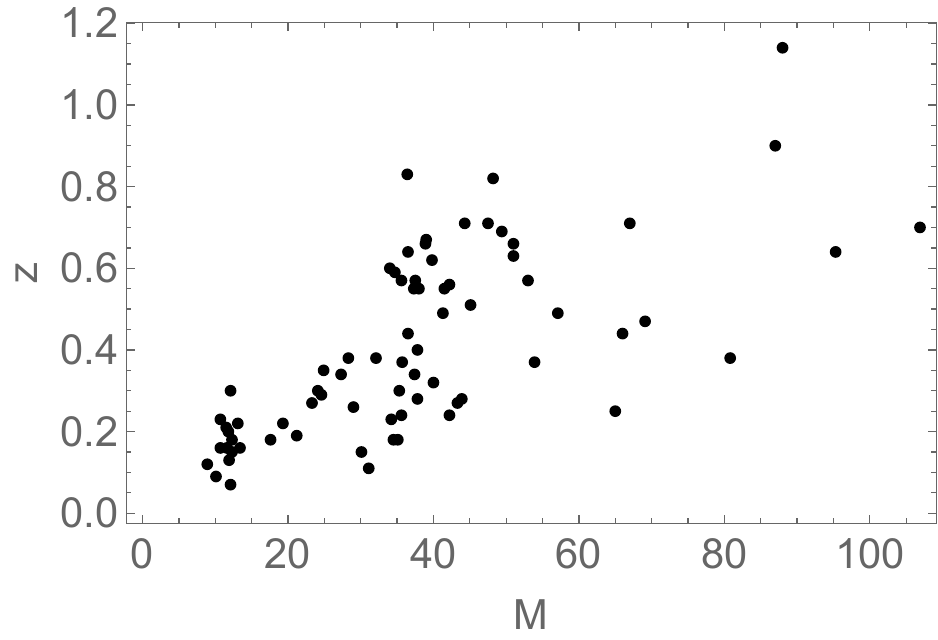}
    \caption{Scatter plot of GW events obtained during the O3 run. Data taken from \cite{LIGOScientific:2021usb, LIGOScientific:2021djp}.}
    \label{o3-catalog}
  \end{center}
\end{figure}

LIGO Scientific, Virgo, and KAGRA Collaboration released the GW events taken during
the third observation run (O3) as GWTC-2.1 and GWTC-3 (\cite{LIGOScientific:2021usb, LIGOScientific:2021djp}).
Excluding small-mass compact objects ($< 3~M_\odot$) that are either BHs or neutron stars,
there are $74$ events that we can reasonably identify as BH-BH merger events.
Fig.~\ref{o3-catalog} shows a scatter plot of those events in the $(M,z)$ plane.
At first glance, this number may appear sufficiently large to enable us to draw 
a statistically meaningful conclusion on the basis of our hypothesis testing.
In this subsection, we will show that the detection bias is crucial, and
this reduces the number of events that can be used owing to the decrease of 
the maximum redshift of usable events.

In all of our analyses presented up to this stage,
we have assumed that the detection probability of the merger events
in the region of the parameter space defined in Fig.~\ref{fig2} is unity, namely, the detection probability $p_{\rm det}$ has been taken to be $1$.
This assumption is valid as long as the detection horizon
of the GW detector is so large that there are sufficiently large numbers of 
merger events that are well inside the detection horizon.
This ideal situation may be achieved by using future detectors,
but may not be achieved in observations by current detectors such as LIGO O3.
Since, in the region where $p_{\rm det}$ is smaller than $1$, 
$p_{\rm det}$ depends nontrivially on $(M,q,z)$ and, in particular, takes a non-separable form in general,
the inclusion of merger events in the region where $p_{\rm det}$
is less than $1$ will degrade
the effectiveness of the hypothesis testing.
Thus, in applying our hypothesis testing to the events obtained during the O3 run, we first need
to restrict the range of the $(M,q)$ space in Fig.~\ref{fig2} to the one where the effect of the selection bias is not significant. 
In practice, this restriction is equivalent to the requirement on $z_c$ such that the distribution of $T_{\rm stat}$ when the underlying distribution
takes the separable form (\ref{merger-rate}) retains nearly the normal distribution $N(0,1)$.

\begin{figure}[t]
\begin{center}
\begin{tabular}{cc}
\begin{minipage}[t]{0.5\hsize}
\includegraphics[width=6cm]{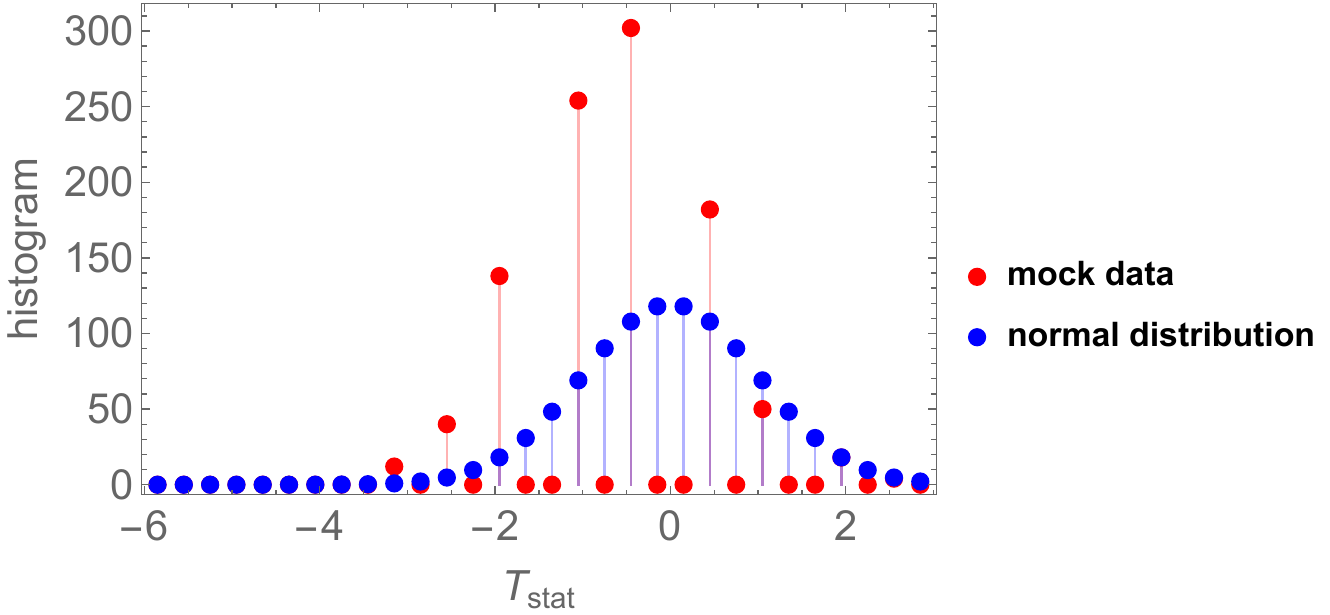}
\end{minipage}
\begin{minipage}[t]{0.5\hsize}
\includegraphics[width=6cm]{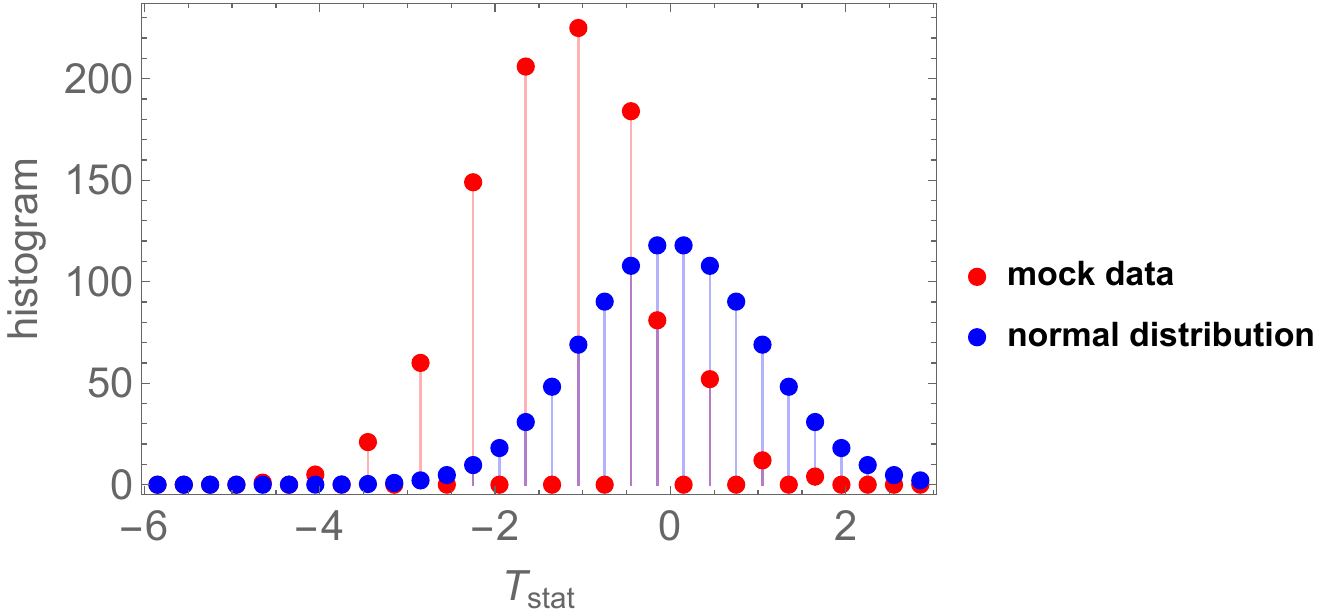}
\end{minipage}
\end{tabular}
\caption{
Left panel: histogram of $T_{\rm stat}$ where $z_c =0.3$ and the selection bias is included.
The number of merger events is taken to be $30$ to be consistent with the O3 catalog. 
Right panel: histogram of $T_{\rm stat}$ where $z_c =0.5$ and the selection bias is included.
The number of merger events is taken to be $48$ to be consistent with the O3 catalog.
}
\label{Tstat-with-selection-bias}
\end{center}
\end{figure}

To investigate such $z_c$ for the LIGO-Virgo network O3 run, we show the histograms of $T_{\rm stat}$
of the mock data obeying the merger rate density of astrophysical BHs used in Sec.~\ref{separable-abh} for two cases $z_c=0.3, 0.5$ in Fig.~\ref{Tstat-with-selection-bias}.
$M_{\rm div}$ dividing regions 1 and 2 has been chosen such that the number of merger events in region 1
becomes equal to that in region 2.
The selection bias has been computed by running a public Python code whose information is given by \cite{Chen:2017wpg}.
To make our analysis consistent with the O3 catalog, we choose $n$, which is the number of merger events,
to be $n=30$ for $z_c=0.3$ and $n=48$ for $z_c =0.5$.
The discontinuous feature of the histograms, which is more prominent in the left panel, is due to the discreteness of $T_{\rm stat}$
caused by the smallness of the sample size $n$.
In the absence of the selection bias, both histograms must obey the normal distribution $N(0,1)$.
We find that the histogram in the right panel ($z_c=0.5$) clearly deviates from the normal distribution.
This suggests that if we apply our method to the O3 catalog by restricting the GW events only to those whose redshift is less than $0.5$,
it can happen with a non-negligible probability that $T_{\rm stat}$ computed from the data lies outside the $2\sigma$ region
even if the mass distribution of the underlying merger rate density does not evolve with the redshifts.
Meanwhile, the histogram in the left panel ($z_c=0.3$) is consistent with the normal distribution.
We expect that the effect of the selection bias is not significant in applying our method to the O3 catalog 
if only the GW events whose redshift is less than $0.3$ are used.

\begin{figure}[t]
  \begin{center}
    \includegraphics[clip,width=6.0cm]{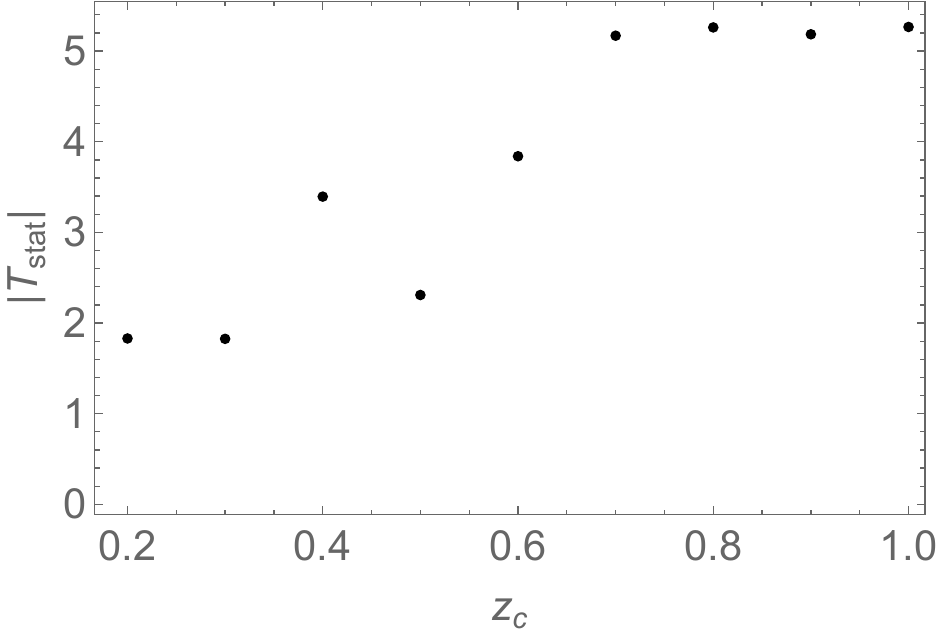}
    \caption{$T_{\rm stat}$ of the O3 catalog for various values of $z_c$ in the range $(0.2, 1,0)$}
    \label{o3-Tstat}
  \end{center}
\end{figure}

Fig.~\ref{o3-Tstat} shows $T_{\rm stat}$ of the O3 catalog for various values of $z_c$ in the range $(0.2, 1,0)$.
For all $z_c$, we find that $T_{\rm stat}$ is negative and the figure shows the absolute value of $T_{\rm stat}$.
From the figure, we observed that although $T_{\rm stat}$ remains within the $2\sigma$ region, it becomes outside 
of the $2\sigma$ region for $z_c \gtrsim 0.4$ and reaches $T_{\rm stat} \simeq -5$ at large $z_c$.
As we have already discussed above, we attribute this behavior to the selection bias. 
Thus, the result that $T_{\rm stat} \simeq -5$ at large $z_c$ does not mean that the O3 data supports that the 
mass distribution of the merger rate evolves with the redshifts.
For $z_c \lesssim 0.3$ where the effect of the selection bias is expected to be unimportant, 
$T_{\rm stat}$ is consistent with the normal distribution.  
On the basis of this observation, we conclude that the current GW observations are consistent with that the 
mass distribution does not evolve with the redshifts.

To summarize, the investigation in this subsection shows that the selection bias degrades the effectiveness
of our method for the O3 catalog by reducing both the number of the merger events and the maximum redshifts ($z_c$).
Within the range where the method can be applied, there is no indication of the time evolution of the mass distribution of the merger rate density.

\section{Conclusions}
There are several known formation channels of the binary BHs that can merge
within the age of the Universe.
However, owing to our lack of theoretical understanding,
we are still far from predicting robustly how much each channel contributes 
to the total merger rate density.
Generally, the fraction of each contribution depends on the BH masses 
as well as the merger redshift.
It is known that some formation channels predict that the time dependence
of the merger rate density is (exactly or nearly) independent of the BH masses.
Naturally, this motivates us to investigate the statistical
testing on the time independence of the mass distribution by which
we may be able to obtain some clues to clarify the origin of the 
binary BHs.
In this paper, we formulated the methodology to perform the above-mentioned test
and demonstrated the effectiveness of the proposed method by using mock data.

After providing the definition of what we exactly mean by 
{\it the mass independence of the time evolution of the merger rate density},
we reformulated it into another equivalent but more convenient form for the statistical analysis.
As a simple statistical test, we adopted the so-called hypothesis testing.
Our null hypothesis is that the time evolution of the 
merger rate density does not depend on the BH masses.
To test the null hypothesis, we introduced the test statistic 
that obeys the normal distribution $N(0,1)$ for the large sample size if the null hypothesis is true.
In Sec.~\ref{demonstration}, by generating the mock data in two specific examples,
both of which satisfy the null hypothesis,
we confirmed explicitly that the test statistic follows the normal distribution.
We also considered two other examples in which the time evolution
of the merger rate density varies for different BH masses and showed that
the central value of the test statistic deviates from zero.
An analytical estimation suggests that the shift of the test statistic is 
proportional to the square of the sample size, and the shift of the 
test statistic computed from the mock data is fairly
consistent with the analytical estimation.
For a given merger rate density that does not fulfill the null hypothesis,
this result supports the reasonable estimate of the minimal sample size
necessary to reject the null hypothesis.
These results demonstrate the effectiveness of our hypothesis testing to
determine from (future) observational data whether the merger
rate density evolves over time independently of the BH masses.

The LIGO-Virgo-KAGRA Collaboration released more than 70 merger events detected during the O3 observation run.
To not undermine the hypothesis testing due to the selection bias 
caused by the low value of the detection probability, 
we investigated how $T_{\rm stat}$ varies as we change the maximum redshift of the merger events that we include
for the computation of $T_{\rm stat}$.
We found that the selection bias degrades the effectiveness of our method for the O3 catalog by reducing both the number of the merger events 
and the maximum redshifts ($z_c$).
Within the range where the method can be applied, the current GW observations are consistent with that the 
mass distribution does not evolve with the redshifts.
This limitation due to the selection bias is expected to be eased in future observations that will deliver much more information
about the merger events in terms of both the number and redshifts.

It should be stressed that our statistical test does not require
a priori specification of the mass distribution, which is largely uncertain, as well as the shape of the time evolution.
Thus, the result of the statistical test is valid independent of the mass distribution and the time evolution.
This is in sharp contrast to previous statistical studies that 
derived/constrained the properties of the BH mergers under specific
assumptions on the mass distribution.

\backmatter

\bmhead{Acknowledgments}

We sincerely thank Bence Kocsis for giving us useful information and suggestions.

\section*{Declarations}

\begin{itemize}
\item Funding \\
This work is supported by JST SPRING, Grant Number JPMJSP2106 (SO). 
This work is also supported by MEXT KAKENHI Grant Number 17H06359~(TS), JP21H05453~(TS) and JSPS KAKENHI Grant Number JP19K03864~(TS).
\item Conflict of interest/Competing interests \\
The authors declare no competing interests.
\item Ethics approval \\
Not applicable.
\item Availability of data and materials \\
The authors confirm that the data supporting the findings of this study are available within the article and the
reference list.
\item Authors' contributions \\
T.~Suyama wrote the main manuscript and S.~Okano prepared the figures and reviewed the manuscript.
\end{itemize}

\bibliography{ref}

\begin{thebibliography}{}
\providecommand{\doi}[1]{\url{https://doi.org/#1}}
\bibcommenthead

\bibitem[\protect\citeauthoryear{Abbott et~al.}{Abbott
  et~al.}{2019}]{LIGOScientific:2018jsj}
Abbott, B.P. et~al. 2019.
\newblock {Binary Black Hole Population Properties Inferred from the First and
  Second Observing Runs of Advanced LIGO and Advanced Virgo}.
\newblock {\em Astrophys. J. Lett.\/}~{\em 882\/}(2): L24.
\newblock \doi{10.3847/2041-8213/ab3800}.
\newblock {\href{https://arxiv.org/abs/1811.12940}{{arXiv:1811.12940}}}
  {[astro-ph.HE]}.

\bibitem[\protect\citeauthoryear{Abbott et~al.}{Abbott
  et~al.}{2021a}]{LIGOScientific:2021usb}
Abbott, R. et~al. 2021a, 8.
\newblock {GWTC-2.1: Deep Extended Catalog of Compact Binary Coalescences
  Observed by LIGO and Virgo During the First Half of the Third Observing Run}.
\newblock {\href{https://arxiv.org/abs/2108.01045}{{arXiv:2108.01045}}}
  {[gr-qc]}.

\bibitem[\protect\citeauthoryear{Abbott et~al.}{Abbott
  et~al.}{2021b}]{LIGOScientific:2021djp}
Abbott, R. et~al. 2021b, 11.
\newblock {GWTC-3: Compact Binary Coalescences Observed by LIGO and Virgo
  During the Second Part of the Third Observing Run}.
\newblock {\href{https://arxiv.org/abs/2111.03606}{{arXiv:2111.03606}}}
  {[gr-qc]}.

\bibitem[\protect\citeauthoryear{Abbott et~al.}{Abbott
  et~al.}{2021c}]{LIGOScientific:2020kqk}
Abbott, R. et~al. 2021c.
\newblock {Population Properties of Compact Objects from the Second LIGO-Virgo
  Gravitational-Wave Transient Catalog}.
\newblock {\em Astrophys. J. Lett.\/}~{\em 913\/}(1): L7.
\newblock \doi{10.3847/2041-8213/abe949}.
\newblock {\href{https://arxiv.org/abs/2010.14533}{{arXiv:2010.14533}}}
  {[astro-ph.HE]}.

\bibitem[\protect\citeauthoryear{Abbott et~al.}{Abbott
  et~al.}{2021d}]{LIGOScientific:2021psn}
Abbott, R. et~al. 2021d, 11.
\newblock {The population of merging compact binaries inferred using
  gravitational waves through GWTC-3}.
\newblock {\href{https://arxiv.org/abs/2111.03634}{{arXiv:2111.03634}}}
  {[astro-ph.HE]}.

\bibitem[\protect\citeauthoryear{Antonini and Perets}{Antonini and
  Perets}{2012}]{Antonini:2012ad}
Antonini, F. and H.B. Perets. 2012.
\newblock {Secular evolution of compact binaries near massive black holes:
  Gravitational wave sources and other exotica}.
\newblock {\em Astrophys. J.\/}~757: 27.
\newblock \doi{10.1088/0004-637X/757/1/27}.
\newblock {\href{https://arxiv.org/abs/1203.2938}{{arXiv:1203.2938}}}
  {[astro-ph.GA]}.

\bibitem[\protect\citeauthoryear{Banerjee}{Banerjee}{2018}]{Banerjee:2017mgr}
Banerjee, S. 2018.
\newblock {Stellar-mass black holes in young massive and open stellar clusters
  and their role in gravitational-wave generation \textendash{} II}.
\newblock {\em Mon. Not. Roy. Astron. Soc.\/}~{\em 473\/}(1): 909--926.
\newblock \doi{10.1093/mnras/stx2347}.
\newblock {\href{https://arxiv.org/abs/1707.00922}{{arXiv:1707.00922}}}
  {[astro-ph.HE]}.

\bibitem[\protect\citeauthoryear{Belczynski et~al.}{Belczynski
  et~al.}{2016}]{Belczynski:2016jno}
Belczynski, K. et~al. 2016.
\newblock {The Effect of Pair-Instability Mass Loss on Black Hole Mergers}.
\newblock {\em Astron. Astrophys.\/}~594: A97.
\newblock \doi{10.1051/0004-6361/201628980}.
\newblock {\href{https://arxiv.org/abs/1607.03116}{{arXiv:1607.03116}}}
  {[astro-ph.HE]}.

\bibitem[\protect\citeauthoryear{Belczynski, Romagnolo, Olejak, Klencki,
  Chattopadhyay, Stevenson, Miller, Lasota, and Crowther}{Belczynski
  et~al.}{2021}]{Belczynski:2021zaz}
Belczynski, K., A.~Romagnolo, A.~Olejak, J.~Klencki, D.~Chattopadhyay,
  S.~Stevenson, M.C. Miller, J.P. Lasota, and P.A. Crowther. 2021, 8.
\newblock {The Uncertain Future of Massive Binaries Obscures the Origin of
  LIGO/Virgo Sources}.
\newblock {\href{https://arxiv.org/abs/2108.10885}{{arXiv:2108.10885}}}
  {[astro-ph.HE]}.

\bibitem[\protect\citeauthoryear{Carr, Raidal, Tenkanen, Vaskonen, and
  Veerm\"ae}{Carr et~al.}{2017}]{Carr:2017jsz}
Carr, B., M.~Raidal, T.~Tenkanen, V.~Vaskonen, and H.~Veerm\"ae. 2017.
\newblock {Primordial black hole constraints for extended mass functions}.
\newblock {\em Phys. Rev. D\/}~{\em 96\/}(2): 023514.
\newblock \doi{10.1103/PhysRevD.96.023514}.
\newblock {\href{https://arxiv.org/abs/1705.05567}{{arXiv:1705.05567}}}
  {[astro-ph.CO]}.

\bibitem[\protect\citeauthoryear{Chatterjee, Rodriguez, Kalogera, and
  Rasio}{Chatterjee et~al.}{2017}]{Chatterjee:2016thb}
Chatterjee, S., C.L. Rodriguez, V.~Kalogera, and F.A. Rasio. 2017.
\newblock {Dynamical Formation of Low-Mass Merging Black Hole Binaries like
  GW151226}.
\newblock {\em Astrophys. J. Lett.\/}~{\em 836\/}(2): L26.
\newblock \doi{10.3847/2041-8213/aa5caa}.
\newblock {\href{https://arxiv.org/abs/1609.06689}{{arXiv:1609.06689}}}
  {[astro-ph.GA]}.

\bibitem[\protect\citeauthoryear{Chen, Holz, Miller, Evans, Vitale, and
  Creighton}{Chen et~al.}{2021}]{Chen:2017wpg}
Chen, H.Y., D.E. Holz, J.~Miller, M.~Evans, S.~Vitale, and J.~Creighton. 2021.
\newblock {Distance measures in gravitational-wave astrophysics and cosmology}.
\newblock {\em Class. Quant. Grav.\/}~{\em 38\/}(5): 055010.
\newblock \doi{10.1088/1361-6382/abd594}.
\newblock {\href{https://arxiv.org/abs/1709.08079}{{arXiv:1709.08079}}}
  {[astro-ph.CO]}.

\bibitem[\protect\citeauthoryear{Dominik, Belczynski, Fryer, Holz, Berti,
  Bulik, Mandel, and O'Shaughnessy}{Dominik et~al.}{2013}]{Dominik:2013tma}
Dominik, M., K.~Belczynski, C.~Fryer, D.E. Holz, E.~Berti, T.~Bulik, I.~Mandel,
  and R.~O'Shaughnessy. 2013.
\newblock {Double Compact Objects II: Cosmological Merger Rates}.
\newblock {\em Astrophys. J.\/}~779: 72.
\newblock \doi{10.1088/0004-637X/779/1/72}.
\newblock {\href{https://arxiv.org/abs/1308.1546}{{arXiv:1308.1546}}}
  {[astro-ph.HE]}.

\bibitem[\protect\citeauthoryear{Fishbach, Doctor, Callister, Edelman, Ye,
  Essick, Farr, Farr, and Holz}{Fishbach et~al.}{2021}]{Fishbach:2021yvy}
Fishbach, M., Z.~Doctor, T.~Callister, B.~Edelman, J.~Ye, R.~Essick, W.M. Farr,
  B.~Farr, and D.E. Holz. 2021.
\newblock {When Are LIGO/Virgo\textquoteright{}s Big Black Hole Mergers?}
\newblock {\em Astrophys. J.\/}~{\em 912\/}(2): 98.
\newblock \doi{10.3847/1538-4357/abee11}.
\newblock {\href{https://arxiv.org/abs/2101.07699}{{arXiv:2101.07699}}}
  {[astro-ph.HE]}.

\bibitem[\protect\citeauthoryear{Fishbach, Holz, and Farr}{Fishbach
  et~al.}{2018}]{Fishbach:2018edt}
Fishbach, M., D.E. Holz, and W.M. Farr. 2018.
\newblock {Does the Black Hole Merger Rate Evolve with Redshift?}
\newblock {\em Astrophys. J. Lett.\/}~{\em 863\/}(2): L41.
\newblock \doi{10.3847/2041-8213/aad800}.
\newblock {\href{https://arxiv.org/abs/1805.10270}{{arXiv:1805.10270}}}
  {[astro-ph.HE]}.

\bibitem[\protect\citeauthoryear{Fragione and Kocsis}{Fragione and
  Kocsis}{2018}]{Fragione:2018vty}
Fragione, G. and B.~Kocsis. 2018.
\newblock {Black hole mergers from an evolving population of globular
  clusters}.
\newblock {\em Phys. Rev. Lett.\/}~{\em 121\/}(16): 161103.
\newblock \doi{10.1103/PhysRevLett.121.161103}.
\newblock {\href{https://arxiv.org/abs/1806.02351}{{arXiv:1806.02351}}}
  {[astro-ph.GA]}.

\bibitem[\protect\citeauthoryear{Fragione and Kocsis}{Fragione and
  Kocsis}{2019}]{Fragione:2019hqt}
Fragione, G. and B.~Kocsis. 2019.
\newblock {Black hole mergers from quadruples}.
\newblock {\em Mon. Not. Roy. Astron. Soc.\/}~{\em 486\/}(4): 4781--4789.
\newblock \doi{10.1093/mnras/stz1175}.
\newblock {\href{https://arxiv.org/abs/1903.03112}{{arXiv:1903.03112}}}
  {[astro-ph.GA]}.

\bibitem[\protect\citeauthoryear{Fragione, Kocsis, Rasio, and Silk}{Fragione
  et~al.}{2022}]{Fragione:2021nhb}
Fragione, G., B.~Kocsis, F.A. Rasio, and J.~Silk. 2022.
\newblock {Repeated Mergers, Mass-gap Black Holes, and Formation of
  Intermediate-mass Black Holes in Dense Massive Star Clusters}.
\newblock {\em Astrophys. J.\/}~{\em 927\/}(2): 231.
\newblock \doi{10.3847/1538-4357/ac5026}.
\newblock {\href{https://arxiv.org/abs/2107.04639}{{arXiv:2107.04639}}}
  {[astro-ph.GA]}.

\bibitem[\protect\citeauthoryear{Fragione, Loeb, and Rasio}{Fragione
  et~al.}{2020}]{Fragione:2020aki}
Fragione, G., A.~Loeb, and F.A. Rasio. 2020.
\newblock {Merging Black Holes in the Low-mass and High-mass Gaps from 2 + 2
  Quadruple Systems}.
\newblock {\em Astrophys. J. Lett.\/}~{\em 895\/}(1): L15.
\newblock \doi{10.3847/2041-8213/ab9093}.
\newblock {\href{https://arxiv.org/abs/2002.11278}{{arXiv:2002.11278}}}
  {[astro-ph.GA]}.

\bibitem[\protect\citeauthoryear{Franciolini, Baibhav, De~Luca, Ng, Wong,
  Berti, Pani, Riotto, and Vitale}{Franciolini
  et~al.}{2021}]{Franciolini:2021tla}
Franciolini, G., V.~Baibhav, V.~De~Luca, K.K.Y. Ng, K.W.K. Wong, E.~Berti,
  P.~Pani, A.~Riotto, and S.~Vitale. 2021, 5.
\newblock {Quantifying the evidence for primordial black holes in LIGO/Virgo
  gravitational-wave data}.
\newblock {\href{https://arxiv.org/abs/2105.03349}{{arXiv:2105.03349}}}
  {[gr-qc]}.

\bibitem[\protect\citeauthoryear{Gond\'an and Kocsis}{Gond\'an and
  Kocsis}{2021}]{Gondan:2020svr}
Gond\'an, L. and B.~Kocsis. 2021.
\newblock {High eccentricities and high masses characterize gravitational-wave
  captures in galactic nuclei as seen by Earth-based detectors}.
\newblock {\em Mon. Not. Roy. Astron. Soc.\/}~{\em 506\/}(2): 1665--1696.
\newblock \doi{10.1093/mnras/stab1722}.
\newblock {\href{https://arxiv.org/abs/2011.02507}{{arXiv:2011.02507}}}
  {[astro-ph.HE]}.

\bibitem[\protect\citeauthoryear{Gond\'an, Kocsis, Raffai, and Frei}{Gond\'an
  et~al.}{2018}]{Gondan:2017wzd}
Gond\'an, L., B.~Kocsis, P.~Raffai, and Z.~Frei. 2018.
\newblock {Eccentric Black Hole Gravitational-Wave Capture Sources in Galactic
  Nuclei: Distribution of Binary Parameters}.
\newblock {\em Astrophys. J.\/}~{\em 860\/}(1): 5.
\newblock \doi{10.3847/1538-4357/aabfee}.
\newblock {\href{https://arxiv.org/abs/1711.09989}{{arXiv:1711.09989}}}
  {[astro-ph.HE]}.

\bibitem[\protect\citeauthoryear{Ioka, Chiba, Tanaka, and Nakamura}{Ioka
  et~al.}{1998}]{Ioka:1998nz}
Ioka, K., T.~Chiba, T.~Tanaka, and T.~Nakamura. 1998.
\newblock {Black hole binary formation in the expanding universe: Three body
  problem approximation}.
\newblock {\em Phys. Rev. D\/}~58: 063003.
\newblock \doi{10.1103/PhysRevD.58.063003}.
\newblock
  {\href{https://arxiv.org/abs/astro-ph/9807018}{{arXiv:astro-ph/9807018}}} .

\bibitem[\protect\citeauthoryear{Kocsis, Suyama, Tanaka, and Yokoyama}{Kocsis
  et~al.}{2018}]{Kocsis:2017yty}
Kocsis, B., T.~Suyama, T.~Tanaka, and S.~Yokoyama. 2018.
\newblock {Hidden universality in the merger rate distribution in the
  primordial black hole scenario}.
\newblock {\em Astrophys. J.\/}~{\em 854\/}(1): 41.
\newblock \doi{10.3847/1538-4357/aaa7f4}.
\newblock {\href{https://arxiv.org/abs/1709.09007}{{arXiv:1709.09007}}}
  {[astro-ph.CO]}.

\bibitem[\protect\citeauthoryear{Madau and Dickinson}{Madau and
  Dickinson}{2014}]{Madau:2014bja}
Madau, P. and M.~Dickinson. 2014.
\newblock {Cosmic Star Formation History}.
\newblock {\em Ann. Rev. Astron. Astrophys.\/}~52: 415--486.
\newblock \doi{10.1146/annurev-astro-081811-125615}.
\newblock {\href{https://arxiv.org/abs/1403.0007}{{arXiv:1403.0007}}}
  {[astro-ph.CO]}.

\bibitem[\protect\citeauthoryear{Mandel and Broekgaarden}{Mandel and
  Broekgaarden}{2021}]{Mandel:2021smh}
Mandel, I. and F.S. Broekgaarden. 2021, 7.
\newblock {Rates of Compact Object Coalescences}.
\newblock {\href{https://arxiv.org/abs/2107.14239}{{arXiv:2107.14239}}}
  {[astro-ph.HE]}.

\bibitem[\protect\citeauthoryear{Mandel, Farr, and Gair}{Mandel
  et~al.}{2019}]{Mandel:2018mve}
Mandel, I., W.M. Farr, and J.R. Gair. 2019.
\newblock {Extracting distribution parameters from multiple uncertain
  observations with selection biases}.
\newblock {\em Mon. Not. Roy. Astron. Soc.\/}~{\em 486\/}(1): 1086--1093.
\newblock \doi{10.1093/mnras/stz896}.
\newblock {\href{https://arxiv.org/abs/1809.02063}{{arXiv:1809.02063}}}
  {[physics.data-an]}.

\bibitem[\protect\citeauthoryear{Mapelli}{Mapelli}{2021}]{Mapelli:2021taw}
Mapelli, M. 2021, 6.
\newblock {Formation channels of single and binary stellar-mass black holes}.
\newblock {\href{https://arxiv.org/abs/2106.00699}{{arXiv:2106.00699}}}
  {[astro-ph.HE]}.

\bibitem[\protect\citeauthoryear{Nakamura, Sasaki, Tanaka, and Thorne}{Nakamura
  et~al.}{1997}]{Nakamura:1997sm}
Nakamura, T., M.~Sasaki, T.~Tanaka, and K.S. Thorne. 1997.
\newblock {Gravitational waves from coalescing black hole MACHO binaries}.
\newblock {\em Astrophys. J. Lett.\/}~487: L139--L142.
\newblock \doi{10.1086/310886}.
\newblock
  {\href{https://arxiv.org/abs/astro-ph/9708060}{{arXiv:astro-ph/9708060}}} .

\bibitem[\protect\citeauthoryear{Raidal, Spethmann, Vaskonen, and
  Veermäe}{Raidal et~al.}{2019}]{Raidal_2019}
Raidal, M., C.~Spethmann, V.~Vaskonen, and H.~Veermäe. 2019, Feb.
\newblock Formation and evolution of primordial black hole binaries in the
  early universe.
\newblock {\em Journal of Cosmology and Astroparticle Physics\/}~{\em
  2019\/}(02): 018–018.
\newblock \doi{10.1088/1475-7516/2019/02/018} .

\bibitem[\protect\citeauthoryear{Rasskazov and Kocsis}{Rasskazov and
  Kocsis}{2019}]{Rasskazov:2019gjw}
Rasskazov, A. and B.~Kocsis. 2019.
\newblock {The rate of stellar mass black hole scattering in galactic nuclei}.
\newblock {\em Astrophys. J.\/}~{\em 881\/}(1): 20.
\newblock \doi{10.3847/1538-4357/ab2c74}.
\newblock {\href{https://arxiv.org/abs/1902.03242}{{arXiv:1902.03242}}}
  {[astro-ph.HE]}.

\bibitem[\protect\citeauthoryear{Rodriguez, Amaro-Seoane, Chatterjee, Kremer,
  Rasio, Samsing, Ye, and Zevin}{Rodriguez et~al.}{2018}]{Rodriguez:2018pss}
Rodriguez, C.L., P.~Amaro-Seoane, S.~Chatterjee, K.~Kremer, F.A. Rasio,
  J.~Samsing, C.S. Ye, and M.~Zevin. 2018.
\newblock {Post-Newtonian Dynamics in Dense Star Clusters: Formation, Masses,
  and Merger Rates of Highly-Eccentric Black Hole Binaries}.
\newblock {\em Phys. Rev. D\/}~{\em 98\/}(12): 123005.
\newblock \doi{10.1103/PhysRevD.98.123005}.
\newblock {\href{https://arxiv.org/abs/1811.04926}{{arXiv:1811.04926}}}
  {[astro-ph.HE]}.

\bibitem[\protect\citeauthoryear{Rodriguez, Chatterjee, and Rasio}{Rodriguez
  et~al.}{2016}]{Rodriguez:2016kxx}
Rodriguez, C.L., S.~Chatterjee, and F.A. Rasio. 2016.
\newblock {Binary Black Hole Mergers from Globular Clusters: Masses, Merger
  Rates, and the Impact of Stellar Evolution}.
\newblock {\em Phys. Rev. D\/}~{\em 93\/}(8): 084029.
\newblock \doi{10.1103/PhysRevD.93.084029}.
\newblock {\href{https://arxiv.org/abs/1602.02444}{{arXiv:1602.02444}}}
  {[astro-ph.HE]}.

\bibitem[\protect\citeauthoryear{Rodriguez and Loeb}{Rodriguez and
  Loeb}{2018}]{Rodriguez:2018rmd}
Rodriguez, C.L. and A.~Loeb. 2018.
\newblock {Redshift Evolution of the Black Hole Merger Rate from Globular
  Clusters}.
\newblock {\em Astrophys. J. Lett.\/}~{\em 866\/}(1): L5.
\newblock \doi{10.3847/2041-8213/aae377}.
\newblock {\href{https://arxiv.org/abs/1809.01152}{{arXiv:1809.01152}}}
  {[astro-ph.HE]}.

\bibitem[\protect\citeauthoryear{Samsing, D'Orazio, Kremer, Rodriguez, and
  Askar}{Samsing et~al.}{2020}]{Samsing:2019dtb}
Samsing, J., D.J. D'Orazio, K.~Kremer, C.L. Rodriguez, and A.~Askar. 2020.
\newblock {Single-single gravitational-wave captures in globular clusters:
  Eccentric deci-Hertz sources observable by DECIGO and Tian-Qin}.
\newblock {\em Phys. Rev. D\/}~{\em 101\/}(12): 123010.
\newblock \doi{10.1103/PhysRevD.101.123010}.
\newblock {\href{https://arxiv.org/abs/1907.11231}{{arXiv:1907.11231}}}
  {[astro-ph.HE]}.

\bibitem[\protect\citeauthoryear{Sasaki, Suyama, Tanaka, and Yokoyama}{Sasaki
  et~al.}{2016}]{Sasaki:2016jop}
Sasaki, M., T.~Suyama, T.~Tanaka, and S.~Yokoyama. 2016.
\newblock {Primordial Black Hole Scenario for the Gravitational-Wave Event
  GW150914}.
\newblock {\em Phys. Rev. Lett.\/}~{\em 117\/}(6): 061101.
\newblock \doi{10.1103/PhysRevLett.117.061101}.
\newblock {\href{https://arxiv.org/abs/1603.08338}{{arXiv:1603.08338}}}
  {[astro-ph.CO]}.

\bibitem[\protect\citeauthoryear{Sasaki, Suyama, Tanaka, and Yokoyama}{Sasaki
  et~al.}{2018}]{Sasaki:2018dmp}
Sasaki, M., T.~Suyama, T.~Tanaka, and S.~Yokoyama. 2018.
\newblock {Primordial black holes\textemdash{}perspectives in gravitational
  wave astronomy}.
\newblock {\em Class. Quant. Grav.\/}~{\em 35\/}(6): 063001.
\newblock \doi{10.1088/1361-6382/aaa7b4}.
\newblock {\href{https://arxiv.org/abs/1801.05235}{{arXiv:1801.05235}}}
  {[astro-ph.CO]}.

\bibitem[\protect\citeauthoryear{Tanikawa, Yoshida, Kinugawa, Trani, Hosokawa,
  Susa, and Omukai}{Tanikawa et~al.}{2021}]{Tanikawa:2021qqi}
Tanikawa, A., T.~Yoshida, T.~Kinugawa, A.A. Trani, T.~Hosokawa, H.~Susa, and
  K.~Omukai. 2021, 10.
\newblock {Merger rate density of binary black holes through isolated
  Population I, II, III and extremely metal-poor binary star evolution}.
\newblock {\href{https://arxiv.org/abs/2110.10846}{{arXiv:2110.10846}}}
  {[astro-ph.HE]}.

\bibitem[\protect\citeauthoryear{Vitale, Farr, Ng, and Rodriguez}{Vitale
  et~al.}{2019}]{Vitale:2018yhm}
Vitale, S., W.M. Farr, K.~Ng, and C.L. Rodriguez. 2019.
\newblock {Measuring the star formation rate with gravitational waves from
  binary black holes}.
\newblock {\em Astrophys. J. Lett.\/}~{\em 886\/}(1): L1.
\newblock \doi{10.3847/2041-8213/ab50c0}.
\newblock {\href{https://arxiv.org/abs/1808.00901}{{arXiv:1808.00901}}}
  {[astro-ph.HE]}.

\bibitem[\protect\citeauthoryear{Yang, Bartos, Haiman, Kocsis, M\'arka, and
  Tagawa}{Yang et~al.}{2020}]{Yang:2020lhq}
Yang, Y., I.~Bartos, Z.~Haiman, B.~Kocsis, S.~M\'arka, and H.~Tagawa. 2020.
\newblock {Cosmic Evolution of Stellar-mass Black Hole Merger Rate in Active
  Galactic Nuclei}.
\newblock {\em Astrophys. J.\/}~{\em 896\/}(2): 138.
\newblock \doi{10.3847/1538-4357/ab91b4}.
\newblock {\href{https://arxiv.org/abs/2003.08564}{{arXiv:2003.08564}}}
  {[astro-ph.HE]}.

\end{thebibliography}


\end{document}